\documentclass{aa}  

\usepackage{graphicx,hyperref}

\usepackage{txfonts}
\newcommand{\dstuc}{DS~Tuc} \newcommand{\xmm}{{\it XMM-Newton}} \newcommand{\pn}{{\it pn}}
\newcommand{\fxu}{erg s$^{-1}$ cm$^{-2}$} \newcommand{\lxu}{erg s$^{-1}$}

\begin{document} \title{X-ray flares of the young planet host \dstuc~A.} 
\subtitle{}

\author{
     I. Pillitteri\inst{1} \and C. Argiroffi\inst{2,1}  \and A. Maggio\inst{1}  
\and G. Micela\inst{1}     \and S. Benatti\inst{1}      \and F. Reale\inst{2}  
\and S. Colombo\inst{1}         \and S.J. Wolk\inst{3} }

\institute{INAF-Osservatorio Astronomico di Palermo, Piazza del Parlamento 1, 90134 Palermo, Italy\\\email{ignazio.pillitteri@inaf.it} 
\and Universit\`a degli Studi di Palermo, Piazza Marina 61, 90133 Palermo, Italy 
\and Harvard-Smithsonian Center for Astrophysics, 02138 Garden St, Cambridge, MA
}

   \date{Received ; accepted }

\abstract{ The discovery of planets around young stars has increased the study of the early phases of
planetary formation and evolution. Stars are strong emitters at X-ray and UV wavelengths in the first
billion of years and this strongly affects the evaporation, thermodynamics and 
chemistry of the atmospheres of young planets around them.  
In order to investigate these effects in young exoplanets we observed the 40 Myr old star \dstuc~A  
with \xmm\ and recorded two X-ray bright flares, with the second event 
occurring about 12 ks after the first one. 
Their duration from the rise to the end of the decay was of about $8-10$ ks in soft X-rays (0.3-10 keV).  The
flares were also recorded in the band 200--300 nm with the UVM2 filter of the Optical Monitor. The duration of the
flares in UV was about 3 ks. The observed delay between the peak in the UV band and in X-rays is a probe
of the heating phase followed by the evaporation and increase of density and emission measure of the flaring loop. 
 The coronal plasma temperature at the two flare peaks reached 54--55 MK.  The diagnostics
based on temperatures and time scales of the flares applied to these two events allow us to infer a
loop length of  $5-7\times10^{10}$ cm,  which is about the size of the stellar radius. We also infer
values of electron density at the flare peaks of $2.3-6.5\times10^{11}$ cm$^{-3}$, and a minimum
magnetic field strength of order of 300--500 G needed to confine the plasma.  
The energy released during the flares was of order of $5-8\times 10^{34}$ erg in the band $0.3-10$ keV
and $0.9-2.7 \times10^{33}$ erg in the UV band (200-300 nm).
We speculate that the flares were associated with Coronal Mass Ejections (CMEs) that hit the planet about
3.3 hr after the flares and dramatically increasing the rate of evaporation of the planet. 
From the RGS spectra we retrieved the emission measure distribution and the abundances of coronal metals during the quiescent 
and the flaring states. In agreement with what inferred from time resolved spectroscopy and EPIC spectra,  
also from the analysis of RGS spectra during the flares we infer a high electron density. }
\keywords{Stars: flare -- Stars: coronae -- Stars: activity -- X-rays: stars }

\maketitle
%

\section{Introduction}
X-rays are emitted by solar type stars during their early evolution \citep{Vaiana1981,Pallavicini81,Feigelson99} and are
a powerful proxy for the structure of the stellar coronae and the magnetic fields  that confine  
the hot coronal plasma.
During the first hundred million  years, solar type stars exhibit very high magnetic and coronal
activity, as well as high stellar rotation rates \citep{Pizzolato2003}.  The loss of angular
momentum  during the Main Sequence is accompanied by a decrease of the { X-ray luminosity} 
by a factor of  $10^3$ during the first billion of years \citep[][]{Pallavicini81, Vaiana1981,Favata03}.

The interest for the study of the activity of stars at high energies
has been renewed with the discovery of extrasolar planets. High energy photons and particles can affect  planetary atmospheres and
their habitability. Fluxes at high energies (X-rays and EUV, XUV) 
have a strong role in shaping and transforming 
the primary atmospheres of the planets { and their evaporation}. 
Because of the decrease of stellar activity, 
most of the effects due to XUV irradiation are expected to happen in the first Gyr  
\citep{Penz08,Cecchi-Pestellini2009,Kubyshkina2018a}.  Collectively, the action of the  magnetized stellar
winds and the coronal mass ejections (CMEs) can make the planets lose  a significant fraction of
their atmospheres and change their compositions through photo-chemical processes (see
\citealp{Drake2013} and \citealp{Kay2019} and references therein). 
However, the effects of the sudden and violent release of energy during flares and CMEs onto the planets 
are still poorly understood due to the lack of proper monitoring of the stellar coronae required in order to 
build a reliable distribution of flaring rates vs. age, and an adequate time-resolved modeling of the 
dynamics and chemistry of the planetary gaseous envelopes.
In this context, the study of young stars hosting planets, like \dstuc, in X-ray and UV bands is of high
relevance.

\dstuc\ is composed by a pair of two young stars { at a distance of 44.1 pc from the Sun,} with an 
age of $\sim40$ Myr and spectral types G6 (component
A and host of the planet) and K3 (component B).  The planet around \dstuc~A was independently
discovered in NASA TESS observations (TOI-200) by \citet{Benatti2019} and \citet{Newton2019}.
\dstuc~Ab has a radius of 0.5 R$_\mathrm J$, a mass upper limit of 14.4 M${\oplus}$ 
\citep{Benatti2021} and { orbits its host star in 8.14 days}.
These characteristics suggest a very low density for the planet and that it is prone to strong atmospheric evaporation.
This is one of few very young planets discovered so far.  Its properties can serve to 
test models of dynamical and atmospheric evolution.  
For the purpose of investigating its coronal emission, \dstuc\
has been observed with \xmm\ in two different observations. 
In \citet{Benatti2021} we have presented the analysis and the global properties of the 
coronae of the  system of \dstuc\ and related the 
X-ray emission to the model of atmospheric evaporation during the Main Sequence.

In the optical band TESS has observed \dstuc~A in six sectors allowing \citet{Colombo2022} to
perform a reconstruction of the distribution of the optical flares and their main properties.
They also inferred the energies of the X-ray counterparts of TESS flares through the
scaling laws of \citet{Flaccomio2018} before X-rays observations were available. 
Here, we study two bright flares of \dstuc~A and discuss their
impact on the atmosphere of \dstuc~Ab.  

The paper is structured as follows: Sect. \ref{analysis}
describes the observations and the data analysis, Sect. \ref{results} presents our results and Sect.
\ref{discussion} contains our discussion and conclusions.

\section{XMM-Newton observations and data analysis}\label{analysis} 
\dstuc\ has been observed two times in X-rays with \xmm\ during observations ObsId=0863400901 
(P.I. S. Wolk) and 0864340101 (P.I. A. Maggio) with exposure times of about 40 ks and 30 ks,
respectively. From the first observation \citep{Benatti2021} we  measured X-ray fluxes and
luminosities of the pair of stars of \dstuc~A and B, separated by $\sim5.3\arcsec$
and partially resolved in MOS detectors.
The A component was 1.3 times more luminous in X-rays than \dstuc~B. 
We also inferred a mean coronal temperature and emission
measure of both stars. Some degree of variability in form of flares was observed in \dstuc~B while 
\dstuc~A was quieter. From these we estimated the time scale of the evaporation of 
\dstuc~Ab and the evolution of its mass and radius in the next 5 Gyr.

The second observation of \dstuc\ with \xmm\ was performed on April 11th 2021 (obsId 0864340101) \footnote{A detailed log of the observation is available at:
\url{https://xmmweb.esac.esa.int/cgi-bin/xmmobs/public/obs\_view.tcl?search\_obs\_id=0864340101} }, again with EPIC as main instrument.  We adopted the {\it Medium} filter and the {\em
SmallWindow} mode for both MOS and \pn.  In addition, we acquired simultaneous high resolution spectra
with RGS and a light curve in the band $200-300$ nm with the Optical Monitor (OM) using the UVM2 filter in Fast mode.  For the OM we obtained five continuous exposures 
of 4400 s and one exposure of 3880 s,  for each exposure a light curve with a time 
binning  of 10 s was obtained.

\begin{figure} 
\begin{center} 
\resizebox{\columnwidth}{!}{ \includegraphics{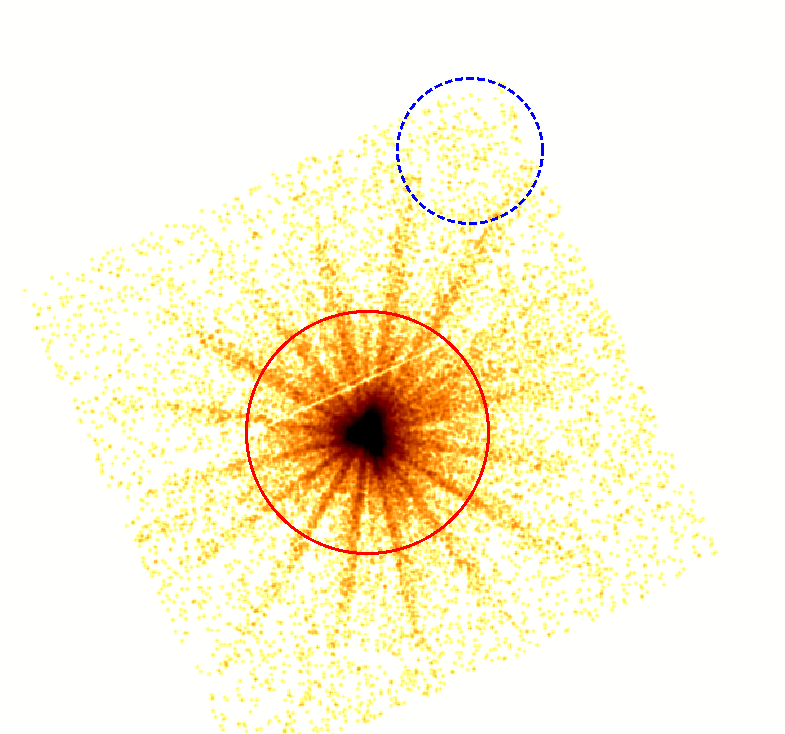} }

\resizebox{\columnwidth}{!}{ \includegraphics{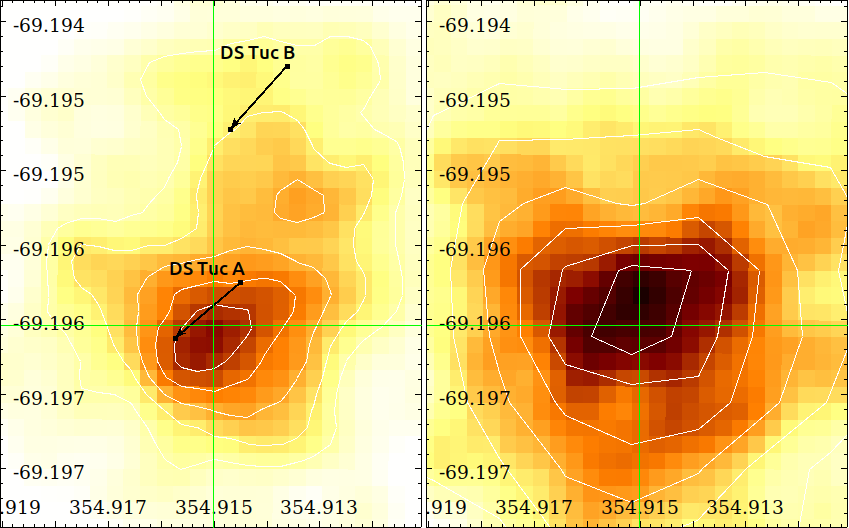} }
\caption{\label{m1_ti1} 
Top panel: MOS 2 image of \dstuc\ with the regions used for accumulating source and
background events. { The scale is logarithmic to enhance the background level
otherwise not visible on a linear scale}. 
The background region was offset to one of the corner of the chip
to avoid the diffraction spikes from the central source.
Bottom left:  an image of MOS 1 during the first 6 ks of
observation and corresponding to interval 1 in Fig. \ref{lc_m1_om}. 
We marked the SIMBAD positions (J2000) of \dstuc~A and B and the vectors of proper motions. 
Bottom right: image during interval 3 corresponding to the peak of the first flare. 
In both images the scale limits and color maps are identical, the binning is 8 pixels corresponding 
to $\sim0.4\arcsec$ and a smoothing with a Gaussian with $\sigma=1.5$ pixels has been applied to images.
We added contour levels with the same ranges for marking the source centroid.} 
\end{center} 
\end{figure}

The ODFs files downloaded from the \xmm\ archive\footnote{\url{https://nxsa.esac.esa.int/nxsa-web/\#home} } were reduced with SAS ver 18.0.0, to 
obtain event tables for EPIC instruments, light curves for OM and spectra for the two RGS
high-resolution spectrometers. We selected the EPIC events for having energies in the band $0.3-10.0$ keV and  {\sc FLAG
== 0} and {\sc PATTERN <= 12} as prescribed by the SAS reduction guide.  We checked that the overall
background was low during the observation, thus the full exposure was retained for further analysis.  A
visual inspection of the \pn\ and MOS images revealed that \dstuc~A underwent strong flaring activity
and its flux dominated the emission of the nearby \dstuc~B. 
Fig. \ref{m1_ti1} shows two MOS1 images relative to two different time intervals and referring
to the quiescent and flare peak intervals of \dstuc~A. 

Spectra and light curves of \dstuc~A were obtained from events in a circular region of radius
75\arcsec centered on the X-ray image centroid (see Fig. \ref{m1_ti1}). 
Spectra of the background were accumulated from an offset
circular region of radius 45\arcsec.  Given the brightness of \dstuc~A in this 
observation we checked for any pile
up that could have affected the spectra of MOS and \pn. We generated the pattern of 
single, double, triple, and quadruple events as a function of the energy as prescribed  by 
the SAS guide.  The diagnostic modeling of the curve did not show evidence of significant 
pile up so we retained the entire list of events in the region. 
 
\begin{figure} 
\begin{center} 
\resizebox{\columnwidth}{!}{ \includegraphics{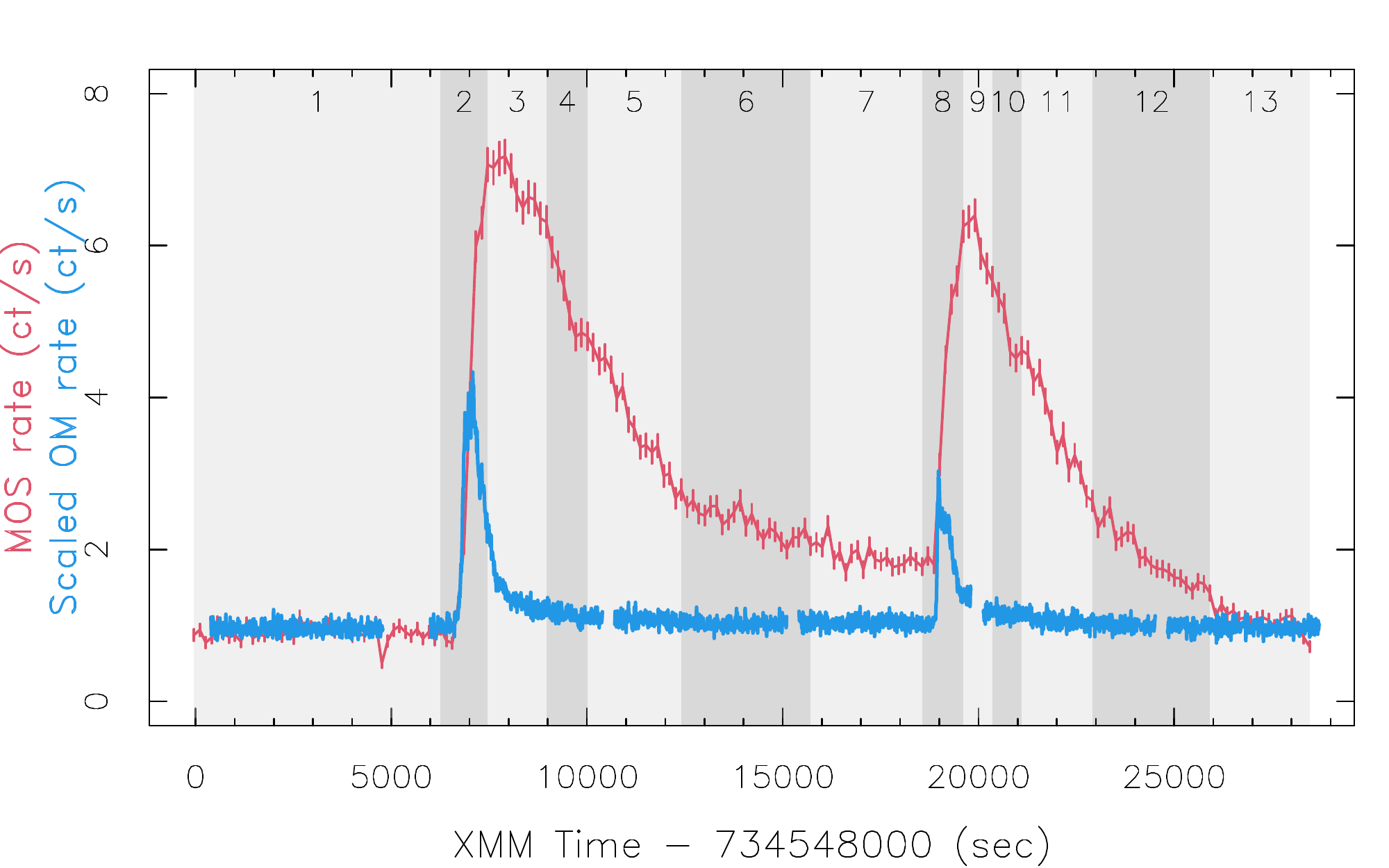} }
\caption{\label{lc_m1_om} Light curves of MOS 1 (red curve, bin size 150 s) 
and OM in UVM2 filter (blue curve, bin size 10 s). OM rate is scaled down by a factor 30
to allow a direct comparison of the rise and decay times of both instruments during
the flares. 
The time intervals used for time resolved spectroscopy (see Sect.
\ref{sect_flares}) are indicated by the alternating light/dark gray areas and 
numbered on top of each panel. } 
\end{center} 
\end{figure}

The high count statistics recorded in the EPIC instruments allowed us to perform time resolved
spectroscopy of the two flares. To this purpose we accumulated the spectra in 13 time intervals that define
 the initial quiescent state, the flare rises, peaks and decays (see Fig. \ref{lc_m1_om}).  
For each time interval the spectra of the source and the background were created with SAS, along 
with the corresponding response matrices and effective area ARF files. 

For the OM we obtained five exposures of 4400 s and one of 3880 s. In each exposure, a light curve 
with time binning of 10 s was accumulated with the standard SAS task {\it omfchain}, and the result is displayed in Fig. \ref{lc_m1_om}.   

For extracting the RGS spectra we used the SAS task {\it rgsproc} and {\it rgscombine} with custom
good time intervals defined as interval nr. 1  for quiescent and 2 to 13 for flares. We summed the
RGS spectrum of the first \xmm\ observation (ObsId 0863400901) to the spectrum of the quiescent interval 
nr. 1, since both are  representative of a quiescent phase or at least with low level of variability. 
Furthermore, we added RGS 1 and RGS 2 spectra of the quiescent and flaring phases for achieving a better count statistics. 

\begin{figure} 
\begin{center} 
\resizebox{\columnwidth}{!}{ \includegraphics{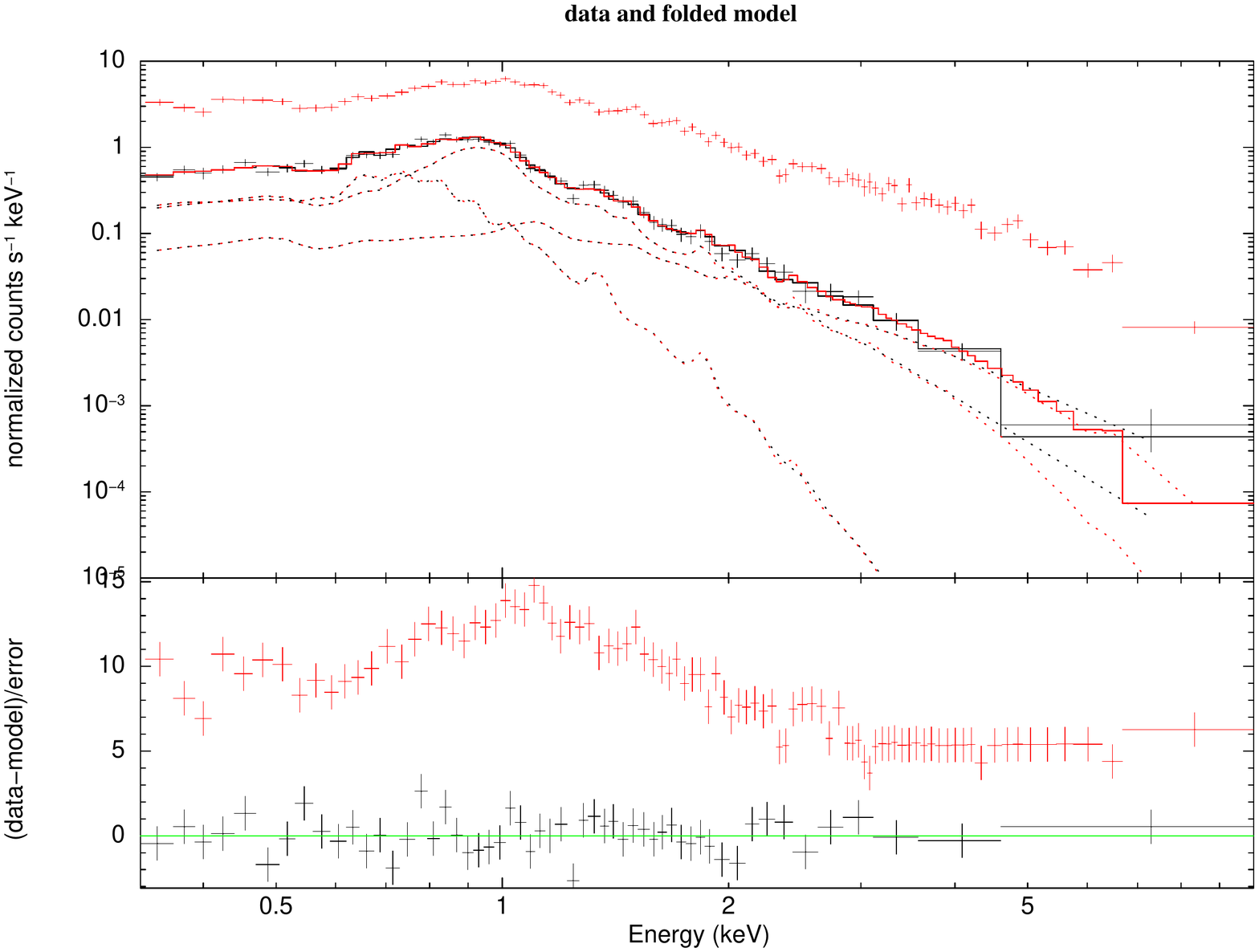} }
\caption{\label{sp_m1_ti1_ti3} Spectra of MOS 1 relative to  the time intervals 1  (black line and crosses) 
and 3 (red crosses).
The best fit model to the spectrum of the plasma during interval 1 with its 3 component is shown along with 
the $\chi$ residuals in the bottom panel. } 
\end{center} 
\end{figure}

\begin{table*} 
\caption{\label{fit_ti1} Best fit parameters of the 3T model for the quiescent spectrum
observed during interval 1. { We list the parameters obtained from the best fit to each EPIC spectrum
and the joint fit of \pn\ and MOS.} Unabsorbed flux and luminosity are calculated in the band 0.3-10.0 keV. }
\begin{center} 
\resizebox{0.7\textwidth}{!}{ 
\begin{tabular}{lrrrr}\hline\hline
Parameter & MOS1 & MOS2 & \pn\ & Joint \\\hline 
N$_\mathrm H$ (10$^{20}$ & 1.5 (0--4.4) & 1.55 (0--4.25) & 2.08 (0.44--3.68) &  2.3 (1.0--3.4)\\
log T$_1$ (K)  & 6.54 (6.48--6.61) & 6.67 (6.6--6.76) & 6.57 (6.52--6.62) & 6.59 (6.56--6.63)\\
log EM1 (cm$^{-3}$) & 52.89 (52.69--53.16) & 52.95 (52.8--53.08) & 52.99 (52.83--53.12) & 53.02 (52.91--53.10) \\
log T$_2$ (K) & 7.03 (6.99--7.05) & 7.06 (7.04--7.07) & 7.01 (7--7.03) & 7.03 (7.02--7.04)\\
log EM2 (cm$^{-3}$) & 52.95 (52.73--53.17) & 53.12 (53.08--53.19) & 53.13 (52.98--53.24) & 53.14 (53.02 53.20)\\
log T$_3$ (K) & 7.42 ($>$7.25) & $>$7.64 & 7.58 ($>$7.34) & 7.87 ($>7.53$) \\
log EM3 (cm$^{-3}$) & 52.51 (51.97--52.73) & 51.97 (51.67--52.15) & 52.32 (51.97--52.6) & 52.08 ($>51.91$)\\
Z/Z$\odot$ & 0.22 (0.13--0.34) & 0.14 (0.11--0.17) & 0.13 (0.0984--0.1804) & 0.13 (0.11--0.16)\\
$\chi^2$  & 52.05 & 55.74 & 60.37  & 190.82 \\
D.o.F.  & 47 & 48 & 48  & 159 \\
P($\chi^2>\chi^2_0$) & 0.28 & 0.21 & 0.11 & 0.04 \\
log f$_\mathrm X$ (\fxu) & -11.11 & -11.13 & -11.1 & -11.09 \\
log L$_\mathrm X$ (\lxu) & 30.26 & 30.23 & 30.27 & 30.27\\
\hline 
\end{tabular} 
} 
\end{center} 
\end{table*}

\begin{table*} \caption{\label{fit_timeseg} Best fit parameters from time resolved spectroscopy of
EPIC spectra.  For each time interval (column 1) are reported the column equivalent absorption
(N$_\mathrm H$), the temperature (kT), the emission measure (E.M.) the global abundance scaling factor
(Z/Z$\odot$), the unabsorbed flux and luminosity in the 0.3-10 keV band, the $\chi^2$, the degrees of
freedom and the probability related to the $\chi^2$ statistics. The parameter ranges refer to the 90\%
confidence intervals.  } 
\resizebox{\textwidth}{!}{ 
\begin{tabular}{lrrrrrrrrr} \\ \hline \hline
Interval & N$_\mathrm H$ & log (T/K) & log E.M.  & Z/Z$\odot$ & log f$_\mathrm X$ & log L$_\mathrm X$  & $\chi^2$ & d.o.f. & Prob \\ 
& 10$^{20}$ cm$^{-2}$&  & cm$^{-3}$ &            & \fxu              & \lxu               &          &        &      \\  
\hline 
& \multicolumn{9}{c}{MOS1}  \\ 
2 & 1.31 (0.17--2.55) & 7.75 (7.67--7.83) & 53.6 (53.57--53.63) & 0.29 (0.25--0.34) & -10.45 & 30.92 & 52.05 & 47 & 0.28 \\
3 & 0.71 (0.06--1.39) & 7.62 (7.59--7.65) & 53.96 (53.94--53.98) & 0.56 (0.5--0.62) & -10.13 & 31.24 & 55.74 & 48 & 0.21 \\
4 & 0 (0--0.58) & 7.51 (7.47--7.54) & 53.84 (53.81--53.87) & 0.47 (0.41--0.53) & -10.29 & 31.08 & 60.37 & 48 & 0.109 \\
5 & 0.45 (0--1.13) & 7.51 (7.48--7.55) & 53.58 (53.56--53.61) & 0.49 (0.45--0.54) & -10.45 & 30.91 & 52.05 & 47 & 0.28 \\
6 & 0.61 (0--1.32) & 7.41 (7.37--7.46) & 53.34 (53.31--53.37) & 0.35 (0.31--0.38) & -10.68 & 30.69 & 55.74 & 48 & 0.21 \\
7 & 0 (0--0.44) & 7.36 (7.32--7.42) & 53.17 (53.13--53.22) & 0.3 (0.27--0.33) & -10.79 & 30.57 & 60.37 & 48 & 0.109 \\
8 & 2.22 (1.02--3.52) & 7.73 (7.67--7.81) & 53.65 (53.62--53.68) & 0.45 (0.39--0.52) & -10.37 & 31 & 52.05 & 47 & 0.28 \\
9 & 1.67 (0.68--2.72) & 7.66 (7.61--7.71) & 53.93 (53.9--53.96) & 0.49 (0.41--0.58) & -10.16 & 31.21 & 55.74 & 48 & 0.21 \\
10 & 0.27 (0--1.49) & 7.61 (7.56--7.67) & 53.77 (53.74--53.8) & 0.51 (0.43--0.59) & -10.29 & 31.07 & 60.37 & 48 & 0.109 \\
11 & 0.83 (0.05--1.66) & 7.53 (7.48--7.57) & 53.6 (53.57--53.63) & 0.44 (0.4--0.49) & -10.46 & 30.91 & 52.05 & 47 & 0.28 \\
12 & 0.15 (0--0.92) & 7.37 (7.32--7.43) & 53.15 (53.1--53.19) & 0.31 (0.28--0.34) & -10.79 & 30.57 & 55.74 & 48 & 0.21 \\
13 & 1.28 (0.11--2.52) & 7.2 (7.09--7.41) & 52.22 (51.82--52.46) & 0.25 (0.22--0.29) & -11.04 & 30.32 & 60.37 & 48 & 0.109 \\ \hline
& \multicolumn{9}{c}{MOS2}  \\ 
2 & 0.65 (0--1.96) & 7.65 (7.58--7.73) & 53.64 (53.59--53.66) & 0.21 (0.17--0.26) & -10.48 & 30.88 & 52.05 & 47 & 0.28 \\
3 & 2.04 (1.33--2.78) & 7.61 (7.58--7.64) & 53.96 (53.94--53.98) & 0.54 (0.48--0.6) & -10.14 & 31.23 & 55.74 & 48 & 0.21 \\
4 & 0.49 (0--1.41) & 7.56 (7.51--7.6) & 53.81 (53.78--53.84) & 0.51 (0.44--0.58) & -10.28 & 31.09 & 60.37 & 48 & 0.109 \\
5 & 1.06 (0.32--1.84) & 7.5 (7.46--7.54) & 53.57 (53.54--53.59) & 0.46 (0.42--0.5) & -10.48 & 30.89 & 52.05 & 47 & 0.28 \\
6 & 1.28 (0.54--2.06) & 7.38 (7.34--7.42) & 53.35 (53.31--53.38) & 0.33 (0.3--0.37) & -10.69 & 30.68 & 55.74 & 48 & 0.21 \\
7 & 0 (0--0.78) & 7.38 (7.34--7.44) & 53.16 (53.11--53.2) & 0.29 (9.27--0.32) & -10.8 & 30.57 & 60.37 & 48 & 0.109 \\
8 & 0.34 (0--1.57) & 7.66 (7.59--7.73) & 53.69 (53.65--53.72) & 0.36 (0.3--0.42) & -10.39 & 30.98 & 52.05 & 47 & 0.28 \\
9 & 1.23 (0.14--2.4) & 7.73 (7.67--7.79) & 53.88 (53.85--53.91) & 0.51 (0.43--0.59) & -10.18 & 31.19 & 55.74 & 48 & 0.21 \\
10 & 1.19 (0--2.48) & 7.57 (7.52--7.63) & 53.8 (53.77--53.84) & 0.4 (0.32--0.49) & -10.31 & 31.06 & 60.37 & 48 & 0.109 \\
11 & 0.5 (0--1.33) & 7.53 (7.49--7.57) & 53.59 (53.56--53.62) & 0.4 (0.35--0.44) & -10.47 & 30.9 & 52.05 & 47 & 0.28 \\
12 & 1.14 (0.27--2.08) & 7.35 (7.29--7.41) & 53.13 (53.07--53.18) & 0.3 (0.27--0.34) & -10.81 & 30.55 & 55.74 & 48 & 0.21 \\
13 & 2.58 (1.34--4.02) & 7.09 (7.02--7.24) & 52.45 (52.08--52.73) & 0.25 (0.19--0.3) & -11.03 & 30.34 & 60.37 & 48 & 0.109 \\ \hline 
& \multicolumn{9}{c}{\pn\ }  \\ 
2 & 0.24 (0--0.81) & 7.7 (7.66--7.75) & 53.72 (53.7--53.74) & 0.18 (0.16--0.2) & -10.37 & 30.99 & 52.05 & 47 & 0.28 \\
3 & 0.16 (0--0.52) & 7.64 (7.62--7.67) & 54.05 (54.04--54.07) & 0.38 (0.36--0.41) & -10.06 & 31.3 & 55.74 & 48 & 0.21 \\
4 & 0.62 (0.15--1.08) & 7.49 (7.47--7.52) & 53.92 (53.9--53.94) & 0.35 (0.32--0.37) & -10.23 & 31.14 & 60.37 & 48 & 0.109 \\
5 & 0.56 (0.22--0.93) & 7.45 (7.43--7.48) & 53.66 (53.64--53.68) & 0.3 (0.29--0.32) & -10.45 & 30.92 & 52.05 & 47 & 0.28 \\
6 & 1.65 (1.28--2.05) & 7.38 (7.35--7.41) & 53.42 (53.4--53.44) & 0.22 (0.21--0.23) & -10.65 & 30.71 & 55.74 & 48 & 0.21 \\
7 & 1.54 (1.12--1.98) & 7.34 (7.3--7.38) & 53.28 (53.25--53.31) & 0.19 (9.18--0.2) & -10.75 & 30.61 & 60.37 & 48 & 0.109 \\
8 & 0.98 (0.43--1.56) & 7.71 (7.66--7.75) & 53.79 (53.77--53.81) & 0.25 (0.23--0.27) & -10.3 & 31.07 & 52.05 & 47 & 0.28 \\
9 & 0.08 (0--0.6) & 7.67 (7.64--7.7) & 54.01 (53.99--54.02) & 0.31 (0.28--0.35) & -10.11 & 31.25 & 55.74 & 48 & 0.21 \\
10 & 0.54 (0--1.14) & 7.56 (7.52--7.6) & 53.86 (53.84--53.89) & 0.31 (0.28--0.34) & -10.26 & 31.1 & 60.37 & 48 & 0.109 \\
11 & 0.25 (0--0.66) & 7.52 (7.49--7.55) & 53.67 (53.65--53.69) & 0.26 (0.24--0.28) & -10.44 & 30.93 & 52.05 & 47 & 0.28 \\
12 & 1.13 (0.72--1.56) & 7.31 (7.28--7.36) & 53.24 (53.21--53.28) & 0.19 (0.17--0.2) & -10.78 & 30.59 & 55.74 & 48 & 0.21 \\
13 & 1.28 (0.72--1.86) & 7.14 (7.05--7.23) & 52.58 (52.43--52.76) & 0.13 (0.11--0.15) & -11.02 & 30.34 & 60.37 & 48 & 0.109 \\
\hline 
\end{tabular} } 
\end{table*}

\begin{table*} 
\caption{\label{btab} Slope in the plane log T vs log EM/2 ($\xi$), loop length, 
electron density and minimum magnetic field derived for the first and second flare 
for each EPIC instrument. Errors in $\xi$ are quoted at 1 $\sigma$ level and propagated to 
the other quantities. Uncertainties in temperatures and decay time $\tau_d$ amount to another 3\%--4\%. } 
\begin{center}\resizebox{0.8\textwidth}{!}{\begin{tabular}{l|ccc} \hline \hline 
Parameter | Instrument         &   MOS1                & MOS2                   & \pn \\ \hline
$\xi_1$ & 0.58 ( 0.47 - 0.68 )  &  0.61 ( 0.52 - 0.69 )  &  0.69 ( 0.56 - 0.82 ) \\
$\xi_2$ & 0.54 ( 0.48 - 0.6 )  &  0.81 ( 0.72 - 0.89 )  &  0.73 ( 0.67 - 0.79 ) \\
L$_1$ (10$^{10}$ cm) & 5.83 ( 3.8 - 7.25 )  &  5.55 ( 4.27 - 6.52 )  &  6.95 ( 5.28 - 8.11 ) \\
L$_2$ (10$^{10}$ cm) & 3.33 ( 2.56 - 3.96 )  &  4.91 ( 4.43 - 8.15 )  &  4.82 ( 4.4 - 5.18 ) \\
n$_{e,1}$ (10$^{11}$ cm) & 2.6 ( 1.9 - 5 )  &  3 ( 2.4 - 4.4 )  &  2.3 ( 1.9 - 3.5 ) \\
n$_{e,2}$ (10$^{11}$ cm) & 6.5 ( 5 - 9.7 )  &  3.8 ( 1.8 - 4.4 )  &  4.4 ( 3.9 - 5 ) \\
B$_1$ (G) & 320 ( 270 - 440 )  &  310 ( 270 - 370 )  &  290 ( 260 - 350 ) \\
B$_2$ (G) & 500 ( 440 - 600 )  &  350 ( 240 - 370 )  &  390 ( 370 - 420 ) \\ \hline 
\end{tabular} }\end{center}
\end{table*}

\section{Results}\label{results}  
We discuss separately the quiescent emission and the flaring emission observed in the present observation.
\subsection{Quiescent phase} \label{sect_quies} The quiescent interval before the first flare was
about 6.5 ks with an average rate of 0.9 ct/s in MOS1, this interval served  to estimate the coronal
plasma properties before the ignition.  We used a sum of 3 APEC thermal components plus a  global
absorption to model the EPIC spectra and obtain a statistically satisfactory description of the
spectrum.  Table \ref{fit_ti1} lists the best fit parameters and Fig. \ref{sp_m1_ti1_ti3} shows the spectrum
during time interval 1 and 3 for comparison. { We performed the best fit procedure to each EPIC spectrum 
and to them jointly obtaining very similar results.}

The main thermal component of the quiescent spectrum is 
 $\log \mathrm{(T/K)} \sim 7.02$, with the addition of a soft component  $\log\mathrm{T/K}\sim0.54$ 
 and a hotter component, much less constrained,  $\log \mathrm{T/K} \ge 7.4$.  
 { By setting a fixed very low value to the equivalent column of gas absorption 
 (N$_\mathrm H = 8\times10^{18}$ cm$^{-2}$)
 we can constrain the hot component at around $\log T \sim 7.4$ (90\% confidence range: 7.31--7.58)
 while the other parameters remain unchanged. However the very low gas absorption is not
 consistent with the values of visual extinction and reddening. }
 
 The temperatures are consistent with the values determined from the analysis of
spectra acquired during the first \xmm\ observation \citep{Benatti2021}. 
The average flux of \dstuc~A in interval 1 is about 1.9 times the flux observed during the first \xmm\ observation.
However, some degree of contamination from \dstuc~B in this estimate has to be accounted for in 
the analysis of this first time interval.  Visual inspection of the image relative to the events of
the quiescent interval shows that \dstuc~A is still the most prominent source and \dstuc~B has a lower
count statistics. The contribution to flux of \dstuc~A from its stellar companion \dstuc~B is of order of 40\% 
during interval 1. { After correcting for this contamination fraction, 
we infer that the quiescent  flux of \dstuc~A in interval 1 was about 15\%  
higher than the average X-ray flux measured during the first \xmm\ observation.} 

\subsection{Flares}\label{sect_flares}
We observed two prominent flares after the initial phase of quiescence that lasted for about 6.5 ks (Fig.\ \ref{lc_m1_om}) 
from the start of the \xmm\ observation. 
The second flare occurred about 12 ks after the beginning  of the first flare.  In Fig. \ref{lc_m1_om} 
we show a light curve of the OM with scaled count rate to allow a straightforward comparison 
of the timing of the flares. 
Each flare starts simultaneously in X-rays and UV, however in the UV band the flares peak earlier 
and decays quicker than in X-ray. 
When the flux reaches each flare peak in UV the rise of the flux in X-rays does flatten while 
continuing to increase. 
The peak in UV corresponds to the end of the heating phase during the flares 
while the loop continues to be filled with plasma evaporating from its feet and thus
increasing the emission measure and the flux in X-rays \citep{Namekata2017}.
This timing and shape of the flares in the near UV band and in X-rays  is analog to the 
Neupert effect seen in the Sun \citep{Neupert1968}.
Neupert noticed that time-integrated microwave fluxes closely match the rising portions of soft X-ray (SXR) emission. 
The effect has been demonstrated since by others (e.g.,  \citealp{Kahler1988}), and was later dubbed as the Neupert effect by 
\citet{Hudson1991}. 
Estimating the time-integrated microwave fluxes from the non-thermal hard X-ray (HXR) emission, 
the Neupert effect has been generalized, 
where the time derivative of the SXR, i.e., $dF_{SHXR}(t)/dt$, can be considered 
as a proxy for the HXR emission $F_{HXR}(t)$  (\citealp{Dennis1993}, \citealp{Veronig2002}). 
In our case, near UV flux measured with OM is a proxy of the $dF_{HXR}/dt$, with HXR 
being the emission in 0.3-10 keV probed by EPIC.

In the UV we measured rise times of $180\pm5$ s and $70\pm5$ s 
for the first  and second flare respectively. 
In X-rays we measured rise times of $900\pm75$ s and $750\pm75$ s, respectively.
The delay between the peak of each flare in UV and X-rays amount to $\sim680$ s 
and $\sim930$ s, respectively. 
In a linear-log plane the decay phase of the flares appear first linear meaning that the first
part of the decay can be modeled with an exponential relationship between count rate and time, then
it does flatten deviating from a pure exponential decay.
With a linear best fit modeling to the first part of the decay of both flares in the linear-log plane
we estimated e-folding times of 3.55 ks  and 2.31 ks for flares 1 and 2, respectively. 
With similar procedure we determined that in UV the flare decays are shorter than in X-rays and
of order of 390 s and 375 s for flares 1 and 2, respectively. 

The evolution of the plasma temperature and its emission measure can provide diagnostics about the
size of the magnetic loop that hosted the flare, the presence of prolonged heating and the minimum
magnetic field required to confine the plasma \citep{Reale2007,Reale2014}.  In order to follow the
evolution of the flares we divided the light curve during the flares in 12 intervals that identify the
rise, the peak and the decay (resolved with 3 different segments each) of each flare as indicated in Fig.
\ref{lc_m1_om}.  For each interval we modeled the extracted spectra with a sum of a 
flaring component described by one APEC optically thin  plasma  thermal emission model added to the  
3-T model that describes the quiescent emission (cf. Sect. \ref{sect_quies}). 
We left free to vary also the global absorption
and the global abundances.  For each time segment the best fit parameters of spectra of MOS 1, 2 and
\pn\ are listed in Table \ref{fit_timeseg}.  { We did not perform a joint best fit to the 3 EPIC spectra
of each time interval because the diagnostics of the flares are calibrated for each instrument separately.}
The absorption is of order of $10^{20}$ cm$^{-2}$, in a
few cases it is not well constrained but we found the 90\% confidence level value always below
$4\times10^{20}$  cm$^{-2}$. These values are in agreement with the estimates of optical extinction
towards \dstuc\ \citep[cf.][]{Benatti2021}. 
\begin{figure*} 
\begin{center} 
\resizebox{\textwidth}{!}{\includegraphics{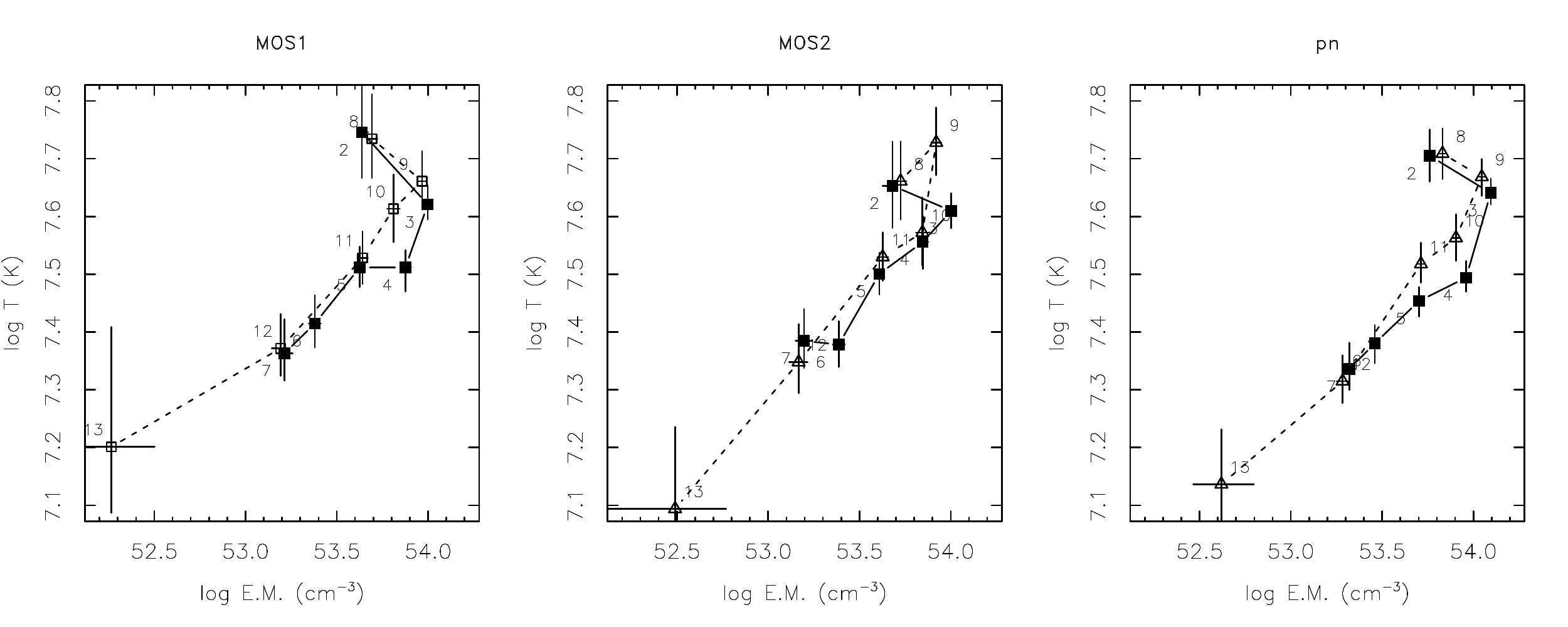}}
\caption{\label{em_logt} Log T (MK) vs. E.M. during the evolution of the two flares.  
Left panel: log T and E.M. derived from MOS1 spectra; central panel: values derived from the MOS2 spectra; 
right panel: parameters derived from \pn\ spectra.  Solid symbols refer to the first flare, open symbols
refer to the second flare.  The numbers refer to the time intervals. The error bars mark the 90\%
confidence interval. } 
\end{center} 
\end{figure*}

Fig. \ref{em_logt} shows the scatter plots of the flare temperatures vs. the emission measures (E.M.)
for each EPIC instrument.  Points 2 to 7 refer to the first flare, points 8 to 13 refer to the second
flare (cf. Table \ref{fit_timeseg}).  
In the spectra of MOS~1 and \pn\ the peak temperatures in both flares are detected in segments 2 
and 8 for flare 1 and 2, respectively, and of order of $\log\mathrm{T/K}\sim7.7-7.75$. 
In MOS~2 for the second flare the peak temperature is detected in segment 9, 
however its value is consistent with that measured in segment 8 with the other two instruments.
The peaks of E.M. are reached in segments 3 and 9 for MOS~1 and \pn\ with values of 
$\log EM(\mathrm{cm}^{-3})\approx54$. 
Owing to our choice of the time intervals, the peak of E.M. in MOS 2
is seen in segment 8 instead of 9, but still consistent with the values determined from MOS 1 and \pn.
We inferred a metallicity value of $Z$ between 0.2 during the decay and 0.5 at the flare peaks. 

The slopes between points 3 and 7, and 9 to 13 in the log T / EM
plane, the temperatures at the peaks of the flares and the e-folding time of the decay of the flares
can be used to infer the size of the loops that hosted the flares using the diagnostics of
\citet{Reale2007} and \citet{Reale2014}. Table \ref{btab} shows the slopes
derived for each flare and EPIC instrument, the total loop lengths, the estimates of the electron
densities and the minimum strength of the magnetic field required to constrain the plasma in the loop
at the peak of both flares.  We inferred a full loop length of order of the stellar radius
($5-7\times10^{10}$ cm).  To estimate the electron density we first calculated the volume of the
flaring loops assumed as tubes with a ratio between base radius and length of 0.1. 
Then we derived the electron densities from E.M., as $n_e = \sqrt{EM/V}$ \citep{Maggio2000}.  Electron
densities at the flare peaks are found of order of $2-6.5\times10^{11}$ cm$^{-3}$.  We can also infer
the minimum strength of the magnetic field needed to confine the plasma as \citep[e.g,][]{Maggio2000}
\[ B=\sqrt{16\pi k_B n_{e,\mathrm{peak}} T_\mathrm{peak}}\]
where $k_B$ is the Boltzmann's constant, $n_{e,\mathrm{peak}}$ and $T_\mathrm{peak}$ are the density
and the temperature at the flare peak. We estimated a minimum magnetic field between 300 G and 500 G.

\subsection{RGS lines} 
\begin{figure}[t] 
\resizebox{0.5\textwidth}{!}{ 
\centering
\includegraphics{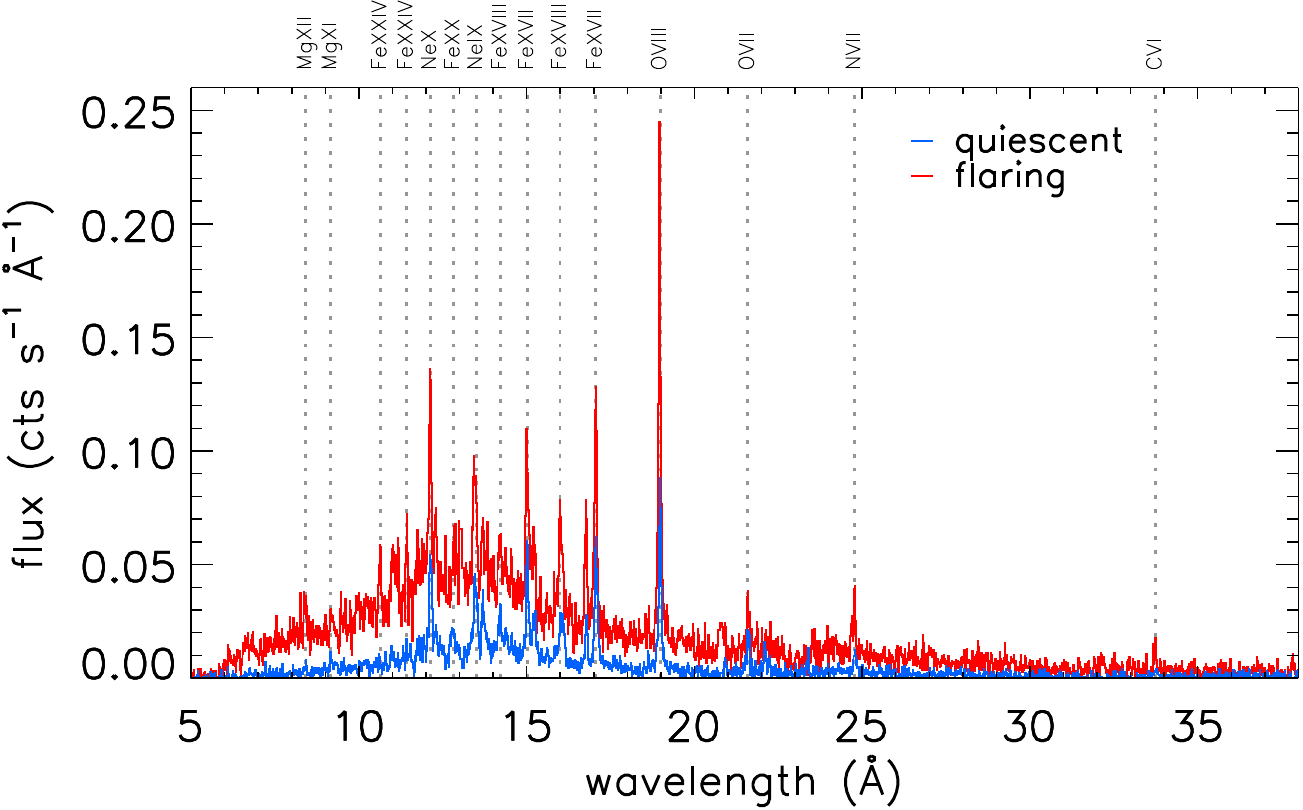} } 
\caption{RGS spectra corresponding to quiescent and flaring phases.  
The identifications of some of the strongest lines are reported on top. 
For clarity reasons the two spectra are slightly smoothed.} 
\label{fig:RGSspec} 
\end{figure}

We analyzed the first-order RGS spectra of both \xmm\ observations to infer the emission measure 
distribution (EMD), abundances, and electron density of the X-ray emitting plasma during the 
quiescent and flaring phases. First order RGS
spectra cover the range $\approx4-38$\,\AA\ with a resolution $0.06$\,\AA\ FWHM. 
The exposures of the quiescent and flaring intervals are 45.53\,ks and 22.15\,ks long, respectively. 
The quiescent spectrum was accumulated summing up the events collected during the whole first \xmm\
observation and the first 6.2 ks of the present observation. For the flaring phase we considered
the remaining 22.2 ks encompassing the two flares altogether.
The RGS spectra of the first order are made of 11200 and 20700 net counts, respectively. 
For the analysis of the RGS data we used the IDL package PINTofALE, \citep{Kashyap2000} 
and the { XSPEC v12.11b} software \citep{Arnaud1996}. 
For the line emissivities we adopted the { CHIANTI atomic database (v7.13, \citealp{Dere1997})}. 

We identified the strongest emission lines and measured their fluxes by fitting the RGS spectra in
small wavelength intervals of $\sim1.0$\,\AA\, width. { The width} of each wavelength interval is set in order to include, and hence fit simultaneously, blended lines. Each observed line was fitted assuming that its
shape is fully described by the RGS line spread function (the intrinsic width of coronal emission
lines is expected to be negligible compared to the RGS line spread function).  
For each wavelength window encompassing the lines to fit we took into account the continuum
contribution in the best fit procedure, leaving temperature and normalization as free parameters. 
Given the small width of each wavelength window, the continuum component from the best fit 
acts as an additive constant value. { This additive constant contribution, mimicking the apparent continuum level, ensures that the contributions of unresolved small lines and wings of other strong lines in the vicinity are taken into account.}
The measured line fluxes are
listed in Table~\ref{tab:RGSfluxes}.  In the case of spectral features suspected to be due to the
blend of more than one line, we listed the different ions contributing to the observed flux.

To determine absolute abundances and EMD values, and to better constrain the hottest 
components of the EMD, continuum flux measurements are also needed. The continuum flux level 
obtained within the line fitting procedures is often an overestimate of the real continuum level, 
because of the significant contribution of small unresolved lines, that is known to affect large 
portions of the RGS wavelength range. Therefore, to obtain reliable measurements of the continuum, 
we selected small wavelength intervals where line contamination is known to be small (on average 
less than 10\%). The continuum flux measurements in these intervals were obtained by integrating the 
observed net counts. The total emissivity function associated to these continuum measurements was 
anyhow computed taking into account also the small fraction of emission line contribution in addition 
to the continuum one. The selected intervals and the corresponding measured fluxes are listed in 
Table~\ref{tab:RGSfluxes}.

\begin{table*}
\caption{Measured fluxes of emission lines and continuum intervals in the X-ray spectrum of DS~Tuc.}
\label{tab:RGSfluxes}
\scriptsize
\begin{center}
\begin{tabular}{@{\hspace{-1mm}}r@{$\div$}l@{\hspace{-1mm}}l@{\hspace{-1mm}}cr@{$\;\pm\;$}l@{\hspace{-3mm}}r@{\extracolsep{4pt}}r@{$\;\pm\;$}l@{\hspace{-3mm}}r@{\hspace{1mm}}c}
\hline\hline
 \multicolumn{2}{c}{ } & & & \multicolumn{3}{c}{quiescent} & \multicolumn{3}{c}{flaring} &  \\
\cline{5-7}\cline{8-10}
\multicolumn{2}{c}{$\lambda^{a}$} & Ion & $\log T_{\rm max}^{b}$ & \multicolumn{2}{c}{$F_{obs}^{c}$} & \multicolumn{1}{c}{$(F_{obs}-F_{pred}^{d})/\sigma$} & \multicolumn{2}{c}{$F_{obs}^{c}$} & \multicolumn{1}{c}{$(F_{obs}-F_{pred}^{d})/\sigma$} &  $EMD^{e}$ \\
\hline
\multicolumn{11}{c}{\it Lines} \\
\hline
\multicolumn{2}{r}{ 8.42\hspace{5mm} } &                                                    {Mg}{XII} {Mg}{XII} &  7.00 &         5.7 &         3.5 &        -1.3\hspace{10mm}~ &        32.4 &        10.1 &        -0.3\hspace{10mm}~ & $\ast$ \\
\multicolumn{2}{r}{ 9.17\hspace{5mm} } &                                                               {Mg}{XI} &  6.80 &        20.0 &         3.9 &         2.6\hspace{10mm}~ &        32.0 &         9.1 &         1.7\hspace{10mm}~ & $\ast$ \\
\multicolumn{2}{r}{ 9.23\hspace{5mm} } &                                                               {Mg}{XI} &  6.80 &         3.2 &         3.2 &         0.6\hspace{10mm}~ &        27.1 &         8.8 &         2.8\hspace{10mm}~ & \\
\multicolumn{2}{r}{ 9.31\hspace{5mm} } &                                                               {Mg}{XI} &  6.80 &        14.2 &         3.4 &         2.9\hspace{10mm}~ &         7.7 &         7.8 &         0.1\hspace{10mm}~ & \\
\multicolumn{2}{r}{10.62\hspace{5mm} } &                   {Fe}{XIX} {Fe}{XIX} {Fe}{XVII} {Fe}{XXIV} {Fe}{XXIV} &  7.30 &         1.6 &         2.6 &        -1.4\hspace{10mm}~ &        51.9 &        12.8 &        -0.0\hspace{10mm}~ & $\ast$ \\
\multicolumn{2}{r}{10.98\hspace{5mm} } &         {Fe}{XXIII} {Ne}{IX} {Na}{X} {Fe}{XXIII} {Fe}{XVII} {Fe}{XXIV} &  7.20 &        22.8 &         4.5 &         3.0\hspace{10mm}~ &        62.1 &        13.8 &         0.4\hspace{10mm}~ & $\ast$ \\
\multicolumn{2}{r}{11.43\hspace{5mm} } &                                      {Fe}{XXII} {Fe}{XXIV} {Fe}{XVIII} &  7.20 &        15.3 &         4.8 &         2.2\hspace{10mm}~ &        41.7 &        13.1 &         1.1\hspace{10mm}~ & $\ast$ \\
\multicolumn{2}{r}{11.74\hspace{5mm} } &                                                            {Fe}{XXIII} &  7.20 &         0.8 &         2.7 &        -2.0\hspace{10mm}~ &        27.1 &        12.6 &        -0.7\hspace{10mm}~ & $\ast$ \\
\multicolumn{2}{r}{11.77\hspace{5mm} } &                                                             {Fe}{XXII} &  7.10 &        13.3 &         4.5 &         1.1\hspace{10mm}~ &        20.8 &        12.5 &        -0.1\hspace{10mm}~ & $\ast$ \\
\multicolumn{2}{r}{12.13\hspace{5mm} } &                                             {Fe}{XVII} {Ne}{X} {Ne}{X} &  6.75 &       106.4 &         8.4 &         1.5\hspace{10mm}~ &       201.8 &        18.6 &        -0.2\hspace{10mm}~ & $\ast$ \\
\multicolumn{2}{r}{12.28\hspace{5mm} } &                                                   {Fe}{XXI} {Fe}{XVII} &  7.05 &        20.9 &         6.0 &        -1.3\hspace{10mm}~ &        54.0 &        14.8 &        -0.0\hspace{10mm}~ & $\ast$ \\
\multicolumn{2}{r}{12.83\hspace{5mm} } &                                             {Fe}{XX} {Fe}{XX} {Fe}{XX} &  7.05 &        27.8 &         5.5 &         1.2\hspace{10mm}~ &        41.6 &        12.8 &         0.2\hspace{10mm}~ & $\ast$ \\
\multicolumn{2}{r}{13.45\hspace{5mm} } &                                 {Fe}{XIX} {Ne}{IX} {Fe}{XIX} {Fe}{XXI} &  6.60 &        57.1 &         7.2 &         0.2\hspace{10mm}~ &       127.2 &        15.6 &         0.5\hspace{10mm}~ & $\ast$ \\
\multicolumn{2}{r}{13.52\hspace{5mm} } &                                                     {Ne}{IX} {Fe}{XIX} &  6.95 &        30.3 &         6.5 &         1.5\hspace{10mm}~ &        34.9 &        13.4 &        -0.4\hspace{10mm}~ & \\
\multicolumn{2}{r}{13.70\hspace{5mm} } &                                                               {Ne}{IX} &  6.55 &        30.5 &         6.2 &         1.7\hspace{10mm}~ &        48.0 &        12.6 &        -0.0\hspace{10mm}~ & \\
\multicolumn{2}{r}{13.74\hspace{5mm} } &                                                              {Fe}{XIX} &  6.95 &        13.1 &         4.7 &         2.3\hspace{10mm}~ &         5.9 &         7.6 &         0.2\hspace{10mm}~ & \\
\multicolumn{2}{r}{13.82\hspace{5mm} } &                                                   {Fe}{XIX} {Fe}{XVII} &  6.90 &        12.3 &         4.0 &        -0.3\hspace{10mm}~ &        35.8 &         9.2 &         1.5\hspace{10mm}~ & \\
\multicolumn{2}{r}{14.20\hspace{5mm} } &                                    {Fe}{XVIII} {Fe}{XVIII} {Fe}{XVIII} &  6.90 &        42.4 &         4.8 &        -0.4\hspace{10mm}~ &        65.0 &         9.7 &        -0.3\hspace{10mm}~ & $\ast$ \\
\multicolumn{2}{r}{14.37\hspace{5mm} } &                                    {Fe}{XVIII} {Fe}{XVIII} {Fe}{XVIII} &  6.90 &        15.0 &         3.8 &         0.0\hspace{10mm}~ &        28.0 &         8.7 &         0.6\hspace{10mm}~ & $\ast$ \\
\multicolumn{2}{r}{14.54\hspace{5mm} } &                                    {Fe}{XVIII} {Fe}{XVIII} {Fe}{XVIII} &  6.90 &         5.6 &         3.5 &        -2.1\hspace{10mm}~ &        23.8 &         8.6 &         0.5\hspace{10mm}~ & $\ast$ \\
\multicolumn{2}{r}{15.01\hspace{5mm} } &                                                             {Fe}{XVII} &  6.75 &        97.9 &         5.9 &         0.9\hspace{10mm}~ &       152.9 &        11.8 &         0.2\hspace{10mm}~ & $\ast$ \\
\multicolumn{2}{r}{15.18\hspace{5mm} } &                                          {Fe}{XIX} {O}{VIII} {O}{VIII} &  6.50 &        19.7 &         4.9 &         1.4\hspace{10mm}~ &        42.5 &        10.6 &         1.2\hspace{10mm}~ & $\ast$ \\
\multicolumn{2}{r}{15.26\hspace{5mm} } &                                                             {Fe}{XVII} &  6.75 &        26.2 &         4.7 &        -0.1\hspace{10mm}~ &         8.2 &         8.7 &        -4.0\hspace{10mm}~ & $\ast$ \\
\multicolumn{2}{r}{16.01\hspace{5mm} } &                                        {Fe}{XVIII} {O}{VIII} {O}{VIII} &  6.50 &        34.1 &         4.5 &        -1.6\hspace{10mm}~ &       100.0 &        10.4 &         0.9\hspace{10mm}~ & $\ast$ \\
\multicolumn{2}{r}{16.07\hspace{5mm} } &                                                            {Fe}{XVIII} &  6.85 &        35.7 &         4.5 &         3.8\hspace{10mm}~ &        40.0 &         9.0 &         1.4\hspace{10mm}~ & $\ast$ \\
\multicolumn{2}{r}{16.78\hspace{5mm} } &                                                             {Fe}{XVII} &  6.75 &        49.5 &         4.6 &        -1.4\hspace{10mm}~ &       109.1 &        10.3 &         1.4\hspace{10mm}~ & $\ast$ \\
\multicolumn{2}{r}{17.05\hspace{5mm} } &                                                             {Fe}{XVII} &  6.75 &        76.3 &         6.6 &         0.7\hspace{10mm}~ &       215.8 &        15.7 &         6.0\hspace{10mm}~ & \\
\multicolumn{2}{r}{17.10\hspace{5mm} } &                                                             {Fe}{XVII} &  6.70 &        59.9 &         7.7 &         0.3\hspace{10mm}~ &        34.5 &        14.0 &        -4.6\hspace{10mm}~ & \\
\multicolumn{2}{r}{17.62\hspace{5mm} } &                                                            {Fe}{XVIII} &  6.90 &         7.9 &         3.0 &        -1.5\hspace{10mm}~ &         8.7 &         6.5 &        -1.6\hspace{10mm}~ & $\ast$ \\
\multicolumn{2}{r}{18.63\hspace{5mm} } &                                                               {O}{VII} &  6.35 &         9.8 &         3.1 &         1.4\hspace{10mm}~ &        17.2 &         7.4 &         1.0\hspace{10mm}~ & $\ast$ \\
\multicolumn{2}{r}{18.97\hspace{5mm} } &                                                    {O}{VIII} {O}{VIII} &  6.50 &       213.9 &         8.8 &         0.2\hspace{10mm}~ &       516.6 &        18.8 &        -0.1\hspace{10mm}~ & $\ast$ \\
\multicolumn{2}{r}{21.60\hspace{5mm} } &                                                               {O}{VII} &  6.30 &        46.4 &         7.1 &         0.5\hspace{10mm}~ &        74.1 &        14.0 &         0.2\hspace{10mm}~ & $\ast$ \\
\multicolumn{2}{r}{21.81\hspace{5mm} } &                                                               {O}{VII} &  6.30 &         0.7 &         3.4 &        -1.8\hspace{10mm}~ &        14.2 &        11.5 &         0.4\hspace{10mm}~ & \\
\multicolumn{2}{r}{22.10\hspace{5mm} } &                                                               {O}{VII} &  6.30 &        37.8 &         6.8 &         2.5\hspace{10mm}~ &        28.8 &        11.5 &        -0.1\hspace{10mm}~ & \\
\multicolumn{2}{r}{24.78\hspace{5mm} } &                                                      {N}{VII} {N}{VII} &  6.30 &        25.4 &         5.3 &        -0.8\hspace{10mm}~ &        72.1 &        12.0 &         0.5\hspace{10mm}~ & $\ast$ \\
\multicolumn{2}{r}{33.73\hspace{5mm} } &                                                        {C}{VI} {C}{VI} &  6.15 &        33.1 &         8.9 &        -0.8\hspace{10mm}~ &        98.4 &        19.2 &         0.5\hspace{10mm}~ & $\ast$ \\
\hline
\multicolumn{11}{c}{\it Continuum} \\
\hline
            $[ 8.49 $ & $  8.90]$ &                                                                        &  7.75 &        25.5 &         4.2 &         0.7\hspace{10mm}~ &       192.1 &        11.6 &        -0.0\hspace{10mm}~ & $\ast$ \\
            $[29.70 $ & $ 30.36]$ &                                                                        &  6.55 &        27.9 &        11.4 &        -0.3\hspace{10mm}~ &       165.3 &        21.3 &         0.1\hspace{10mm}~ & $\ast$ \\
\hline
\end{tabular}
\end{center}
$^a$~Wavelengths (\AA).
$^b$~Temperature (K) of maximum emissivity.
$^c$~Observed fluxes (${\rm 10^{-6}\,ph\,s^{-1}\,cm^{-2}}$) with uncertainties at the 68\% confidence level.
$^d$~Predicted fluxes considered here those computed in the low density limit.
$^e$~Flux measurements selected for the $EMD$ reconstruction.
\normalsize
\end{table*}
\normalsize

\begin{figure}
    \centering
    \resizebox{\columnwidth}{!}{\includegraphics{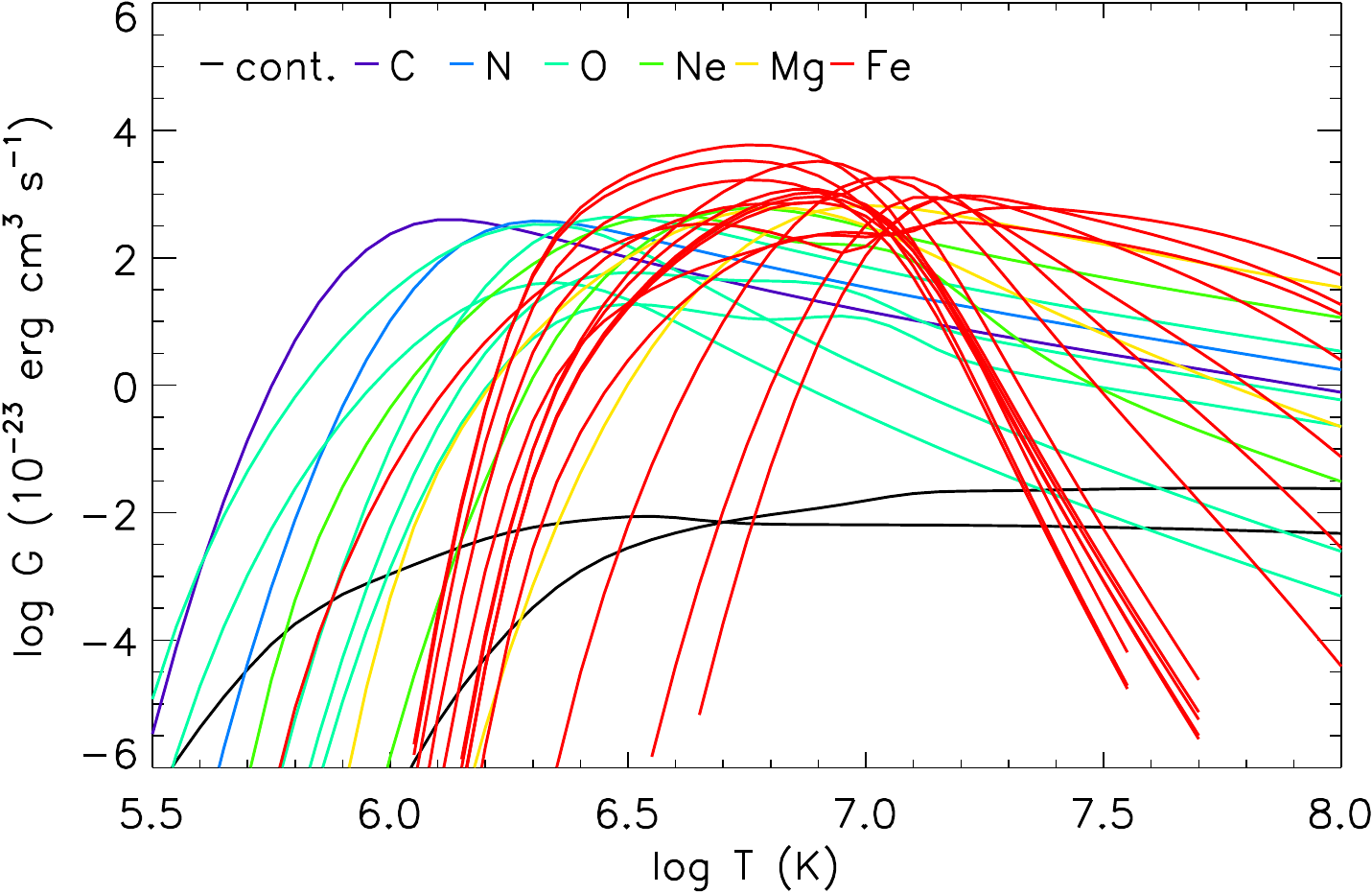}}    
    \caption{Emissivity functions of the ion lines (colored curves) and continuum (black curves)  
    measured in RGS spectra.}
    \label{fig:emis}
\end{figure}

\begin{figure}[t] 
\resizebox{\columnwidth}{!}{ 
\centering \includegraphics{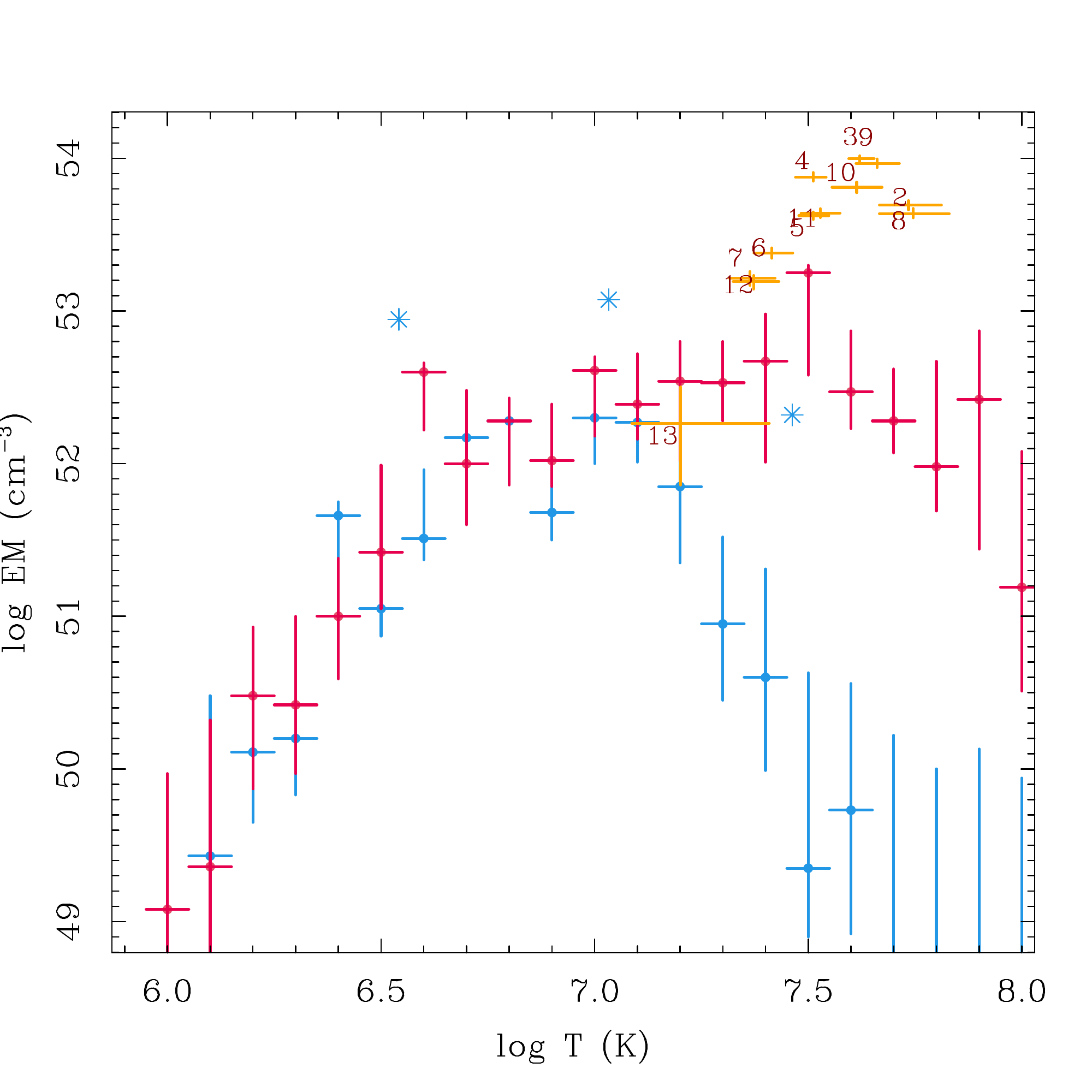} }
\caption{EMDs derived from the RGS line fluxes for the quiescent (blue) and flaring phases (red).
Asterisks and solid points mark the EM derived from EPIC detectors for the quiescent 
(blue symbol, 3T APEC) and flaring intervals, (red symbols and interval numbers, 1T APEC).}
\label{fig:plotemd} \end{figure}

\normalsize
We derived the EMD and the abundances corresponding to the quiescent and flaring phases by applying
the Markov chain Monte Carlo method of \citet{Kashyap1998}. 
Flux measurements used for the EMD reconstruction are listed in Table~\ref{tab:RGSfluxes}.
We discarded density sensitive lines and lines whose fluxes are highly uncertain because 
of strong blending.
Measured fluxes were first converted into unabsorbed fluxes assuming a hydrogen column density
$N_{\mathrm H} = 2\times10^{20}\mathrm{cm^{-2}}$, this is an average value 
inferred from the analysis of EPIC spectra (cf. Sect.\ref{sect_quies}), and from that reported by
\citet{Benatti2021}. 
Given the range of formation temperatures of the measured fluxes, we reconstructed the EMD over a
regular logarithmic temperature grid, with bins of width of 0.10 dex, and encompassing the range of
$\log \mathrm T(\mathrm K)$ from 6.0 to 8.0 dex.
We show in Fig. \ref{fig:emis} the emissivity functions associated with the line and continuum flux 
measurements used for the EMD derivation. 
The hot tails that characterize the shapes of the emissivity functions of several lines, 
as well as that of the continuum intervals, make their associated flux measurements highly sensitive to the 
amount of hot plasma components. 
Moreover, { since we use both line and continuum flux measurements to reconstruct the EMD, the inversion procedure} provides simultaneously the absolute abundances for those elements for
which at least one line has been measured. For the measured spectral features that include lines of
different elements, the total emissivity, that is needed for the EMD reconstruction, depends on the
starting abundance values of the elements involved. For this reason, the EMD and abundance
reconstruction have been performed recursively. The first attempts of EMD reconstruction indicated that
the abundances of the quiescent and flaring phases do not differ significantly from each other.
Because of that, in the final EMD derivation of the quiescent phase we assumed the same set of
abundances obtained from the analysis of the flaring phase, where, due to the higher line fluxes,
they are determined with better precision.
The EMD and the abundances derived for the X-ray emitting plasma of \dstuc~A are reported in
Table~\ref{tab:RGSmodel}. The two EMDs are also plotted in Fig.~\ref{fig:plotemd}.

\begin{table}
\caption{EMD and abundances of DS Tuc.}
\label{tab:RGSmodel}
\small
\begin{center}
\begin{tabular}{c@{\hspace{3mm}}c@{\hspace{3mm}}c}
\hline\hline
          & quiescent & flaring \\
log T (K) & log EMD (cm$^{-3}$) & log EMD (cm$^{-3}$) \\
\hline
6.00 & 48.36$^{+ 1.50}_{- 0.59}$  & 49.08$^{+ 0.89}_{- 1.01}$  \\
6.10 & 49.43$^{+ 1.05}_{- 0.45}$  & 49.36$^{+ 0.96}_{- 0.59}$  \\
6.20 & 50.11$^{+ 0.77}_{- 0.46}$  & 50.48$^{+ 0.45}_{- 0.61}$  \\
6.30 & 50.20$^{+ 0.59}_{- 0.37}$  & 50.42$^{+ 0.58}_{- 0.45}$  \\
6.40 & 51.66$^{+ 0.09}_{- 0.99}$  & 51.00$^{+ 0.38}_{- 0.41}$  \\
6.50 & 51.05$^{+ 0.47}_{- 0.18}$  & 51.42$^{+ 0.57}_{- 0.37}$  \\
6.60 & 51.51$^{+ 0.45}_{- 0.14}$  & 52.60$^{+ 0.06}_{- 0.38}$  \\
6.70 & 52.17$^{+ 0.10}_{- 0.32}$  & 52.00$^{+ 0.48}_{- 0.40}$  \\
6.80 & 52.28$^{+ 0.11}_{- 0.34}$  & 52.28$^{+ 0.15}_{- 0.42}$  \\
6.90 & 51.68$^{+ 0.50}_{- 0.18}$  & 52.02$^{+ 0.37}_{- 0.17}$  \\
7.00 & 52.30$^{+ 0.11}_{- 0.30}$  & 52.61$^{+ 0.09}_{- 0.43}$  \\
7.10 & 52.27$^{+ 0.15}_{- 0.26}$  & 52.39$^{+ 0.33}_{- 0.23}$  \\
7.20 & 51.85$^{+ 0.22}_{- 0.50}$  & 52.54$^{+ 0.26}_{- 0.52}$  \\
7.30 & 50.95$^{+ 0.57}_{- 0.50}$  & 52.53$^{+ 0.27}_{- 0.25}$  \\
7.40 & 50.60$^{+ 0.71}_{- 0.61}$  & 52.67$^{+ 0.31}_{- 0.66}$  \\
7.50 & 49.35$^{+ 1.28}_{- 0.45}$  & 53.25$^{+ 0.05}_{- 0.67}$  \\
7.60 & 49.73$^{+ 0.83}_{- 0.81}$  & 52.47$^{+ 0.40}_{- 0.24}$  \\
7.70 & 48.42$^{+ 1.80}_{- 0.43}$  & 52.28$^{+ 0.34}_{- 0.21}$  \\
7.80 & 47.43$^{+ 2.57}_{- 0.52}$  & 51.98$^{+ 0.69}_{- 0.29}$  \\
7.90 & 47.33$^{+ 2.80}_{- 0.54}$  & 52.42$^{+ 0.45}_{- 0.98}$  \\
8.00 & 47.47$^{+ 2.47}_{- 0.79}$  & 51.19$^{+ 0.89}_{- 0.68}$  \\
\hline
Elem & \multicolumn{2}{c}{$A_X/A_{X\odot}$} \\
\hline
C & \multicolumn{2}{c}{  0.73$^{+ 0.28}_{- 0.16}$} \\ 
N & \multicolumn{2}{c}{  0.85$^{+ 0.35}_{- 0.15}$} \\ 
O & \multicolumn{2}{c}{  0.54$^{+ 0.16}_{- 0.08}$} \\ 
Ne & \multicolumn{2}{c}{ 0.91$^{+ 0.30}_{- 0.16}$} \\ 
Mg & \multicolumn{2}{c}{ 0.41$^{+ 0.19}_{- 0.10}$} \\ 
Fe & \multicolumn{2}{c}{ 0.26$^{+ 0.07}_{- 0.05}$} \\ 
\hline
\end{tabular}
\end{center}
Abundances are in solar units (Anders \& Grevesse 1989, GeCoA, 53, 97).
\normalsize
\end{table}
\normalsize

\normalsize
It is possible to constrain the X-ray emitting plasma density $n_{\mathrm{e}}$ by inspecting the
diagnostics provided by He-like line triplets and other density-sensitive line ratios.  Among the
emission lines detected in the flaring and quiescent RGS spectra, density-sensitive line ratios are
provided by the He-like triplets of Mg~XI (resonance $r$ at 9.17\,\AA, intercombination $i$ at
9.23\,\AA, forbidden $f$ at 9.31\,\AA), Ne~IX ($\lambda_r=13.45$\,\AA, $\lambda_i=13.55$\,\AA,
$\lambda_f=13.70$\,\AA),  and O~VII ($\lambda_r=21.60$\,\AA, $\lambda_i=21.81$\,\AA, and
$\lambda_f=22.10$\,\AA), and by the pair of  Fe~XVII lines at 17.05\AA\ and 17.10\,\AA\ \citep{Mauche2001}.

However, we did not analyzed the He-like Ne~IX triplet at $\sim 13.5$\,\AA, because of its severe blending
with strong Fe lines makes its measured fluxes highly uncertain. 
The O~VII triplet, that has a maximum formation temperature of $\sim2$\,MK, allows us to measure the
density of relatively cold coronal plasma. The Mg~XI and Fe~XVII density diagnostics instead probe
hotter plasma components, since their line maximum formation temperatures are at $\sim6$\,MK.

\begin{figure*}[t] 
\centering 
\includegraphics[scale=0.57]{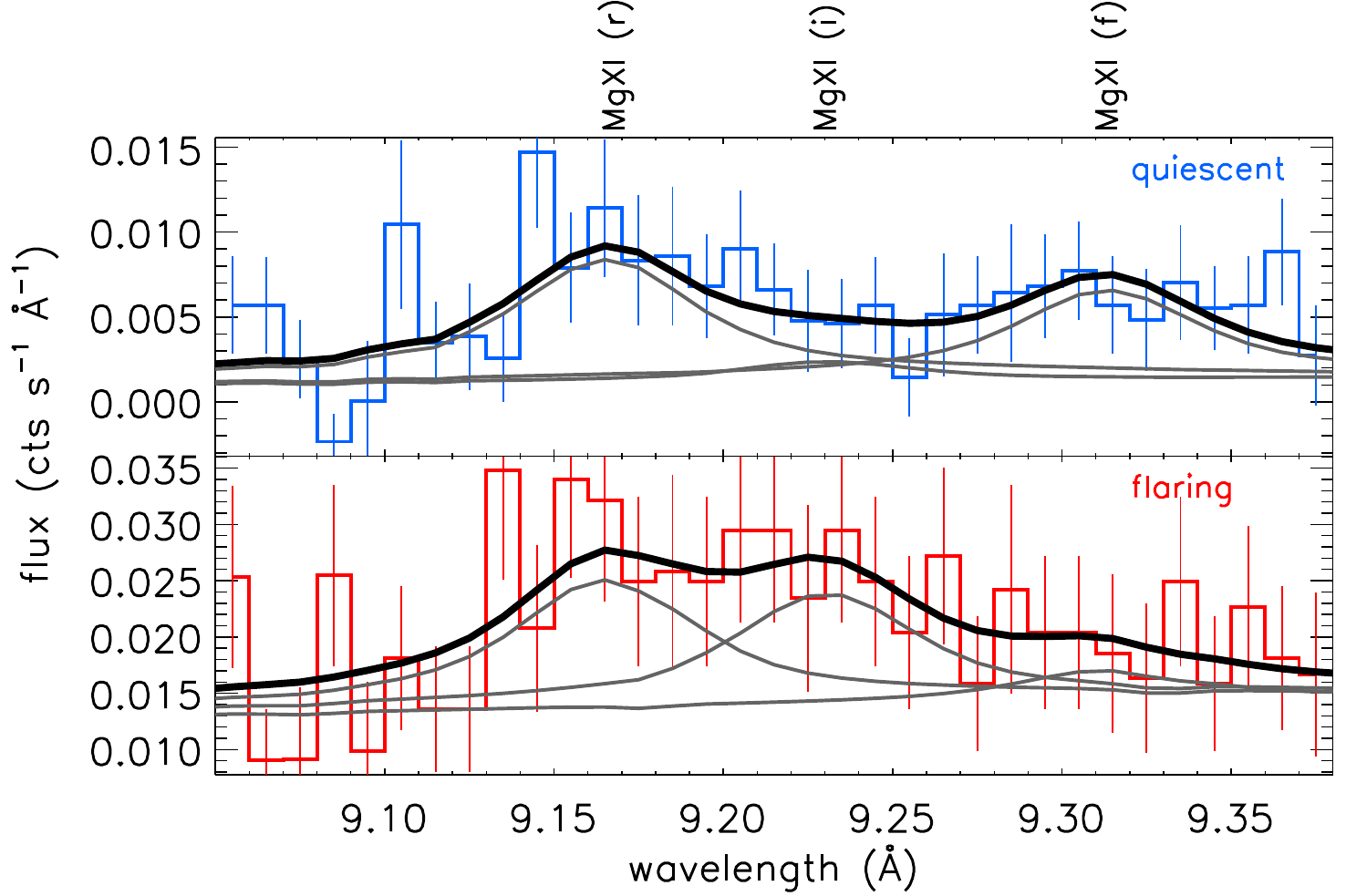}
\includegraphics[scale=0.57]{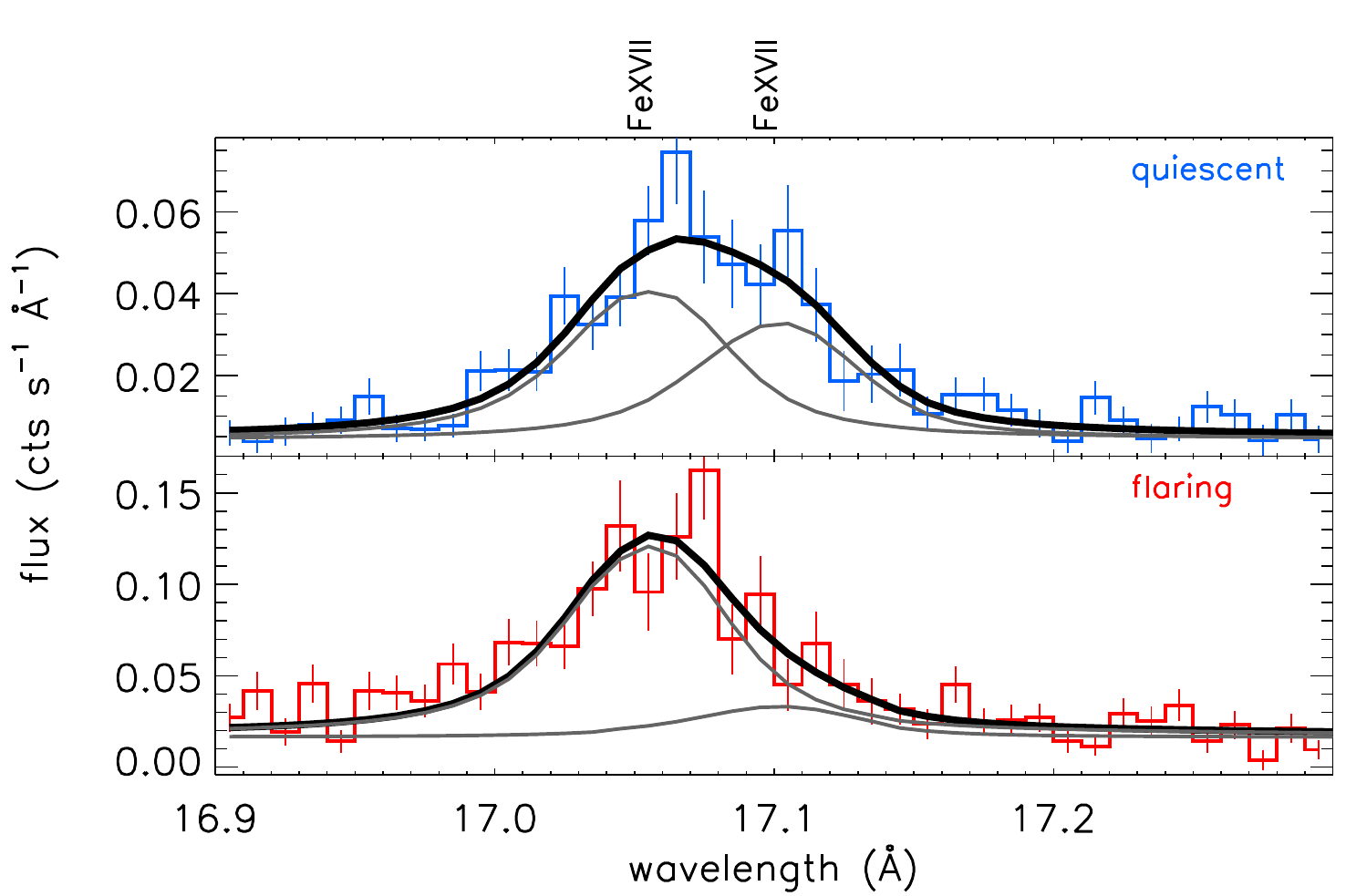} \caption{Observed (histogram with error bars) and
best fit (solid black curve) spectra, of the quiescent and flaring phases, in the regions of the Mg~XI
triplet and Fe~XVII density sensitive lines.  Best fit functions corresponding to individual line are
also shown with solid grey lines.  {\it Left panels}; $r$, $f$, and $i$ labels indicate the resonance,
inter-combination, and forbidden lines of the Mg~XI triplet.} 
\label{fig:densitylines} 
\end{figure*}

The $f/i$ line ratio of the O~VII triplet is high and compatible with the low density limits for both
the quiescent and flaring phases. Density sensitive lines of Mg~XI and Fe~XVII, of the quiescent and
flaring phases, are displayed in Fig.~\ref{fig:densitylines}.  Both the quiescent and flaring spectra
in the Mg~XI region are quite noisy due to the low S/N ratio of the lines, and to the
significant continuum emission in this region, especially during  flares.  The Fe~XVII
lines are well exposed, with a good $S/N$ ratio in both the quiescent and flare spectra. 
However, the Fe~XVII lines are separated by 0.05\,\AA\ which is comparable 
with the RGS spectral resolution (0.06\,\AA) making the two lines 
only marginally resolved.  Because of all these sources of uncertainty, the Mg~XI and Fe~XVII line
fluxes were obtained by freezing line positions, and leaving the line fluxes as free parameters for the
best fit of the line profiles.

\begin{figure}[t] 
\resizebox{\columnwidth}{!}{ \centering \includegraphics{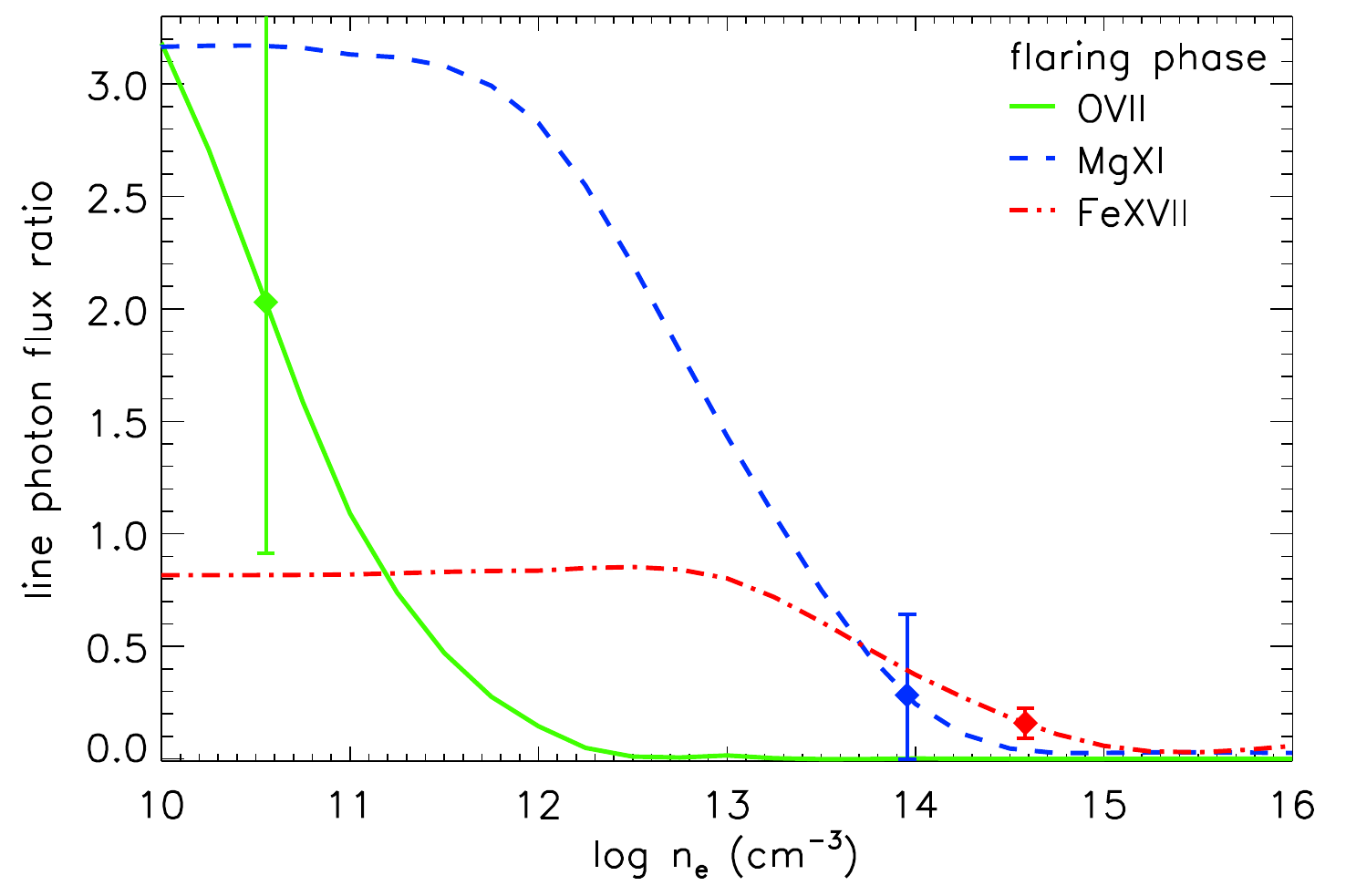} }
\caption{Predicted line ratios ($f/i$ for O~VII and Mg~XI and $F(17.10)/F(17.05)$ for Fe~XVII)
computed at different plasma densities, together with the observed line ratios observed during the
flaring phase.} 
\label{fig:measure_density} 
\end{figure}

The observed $f/i$ ratio of Mg~XI and the $F(17.10)/F(17.05)$ ratio of Fe~XVII change significantly
during the quiescent and flaring phases. During quiescent emission both Mg~XI and Fe~XVII have both
ratios compatible with the low density limit: $f/i>2$ and $F(17.10)/F(17.05)=0.87^{+0.13}_{-0.12}$,
both indicating $n_{\mathrm{e}}<10^{13}$\,cm$^{-3}$. Conversely, during the flaring phases, both
ratios decrease significantly  ($f/i=0.28^{+0.36}_{-0.28}$ and $F(17.10)/F(17.05)=0.16\pm0.07$)
indicating a high plasma density ($n_{\mathrm{e}}\sim10^{14}$\,cm$^{-3}$), as shown in
Fig.~\ref{fig:measure_density}. This value is derived from relatively cold lines and they probe
the cooler plasma in the loop which is toward its feet \citep[][]{Argiroffi2019}. 
Conversely, the plasma density inferred
from the emission measure and the loop volume described in Sect. \ref{sect_flares} is valid 
for hotter plasma towards the apex of the loop during the flare peak.
Also, in deriving the electron density from the flare emission measure, 
the choice of the volume has some degree of arbitrarity. 

\begin{figure}
    \centering
    \resizebox{\columnwidth}{!}{\includegraphics{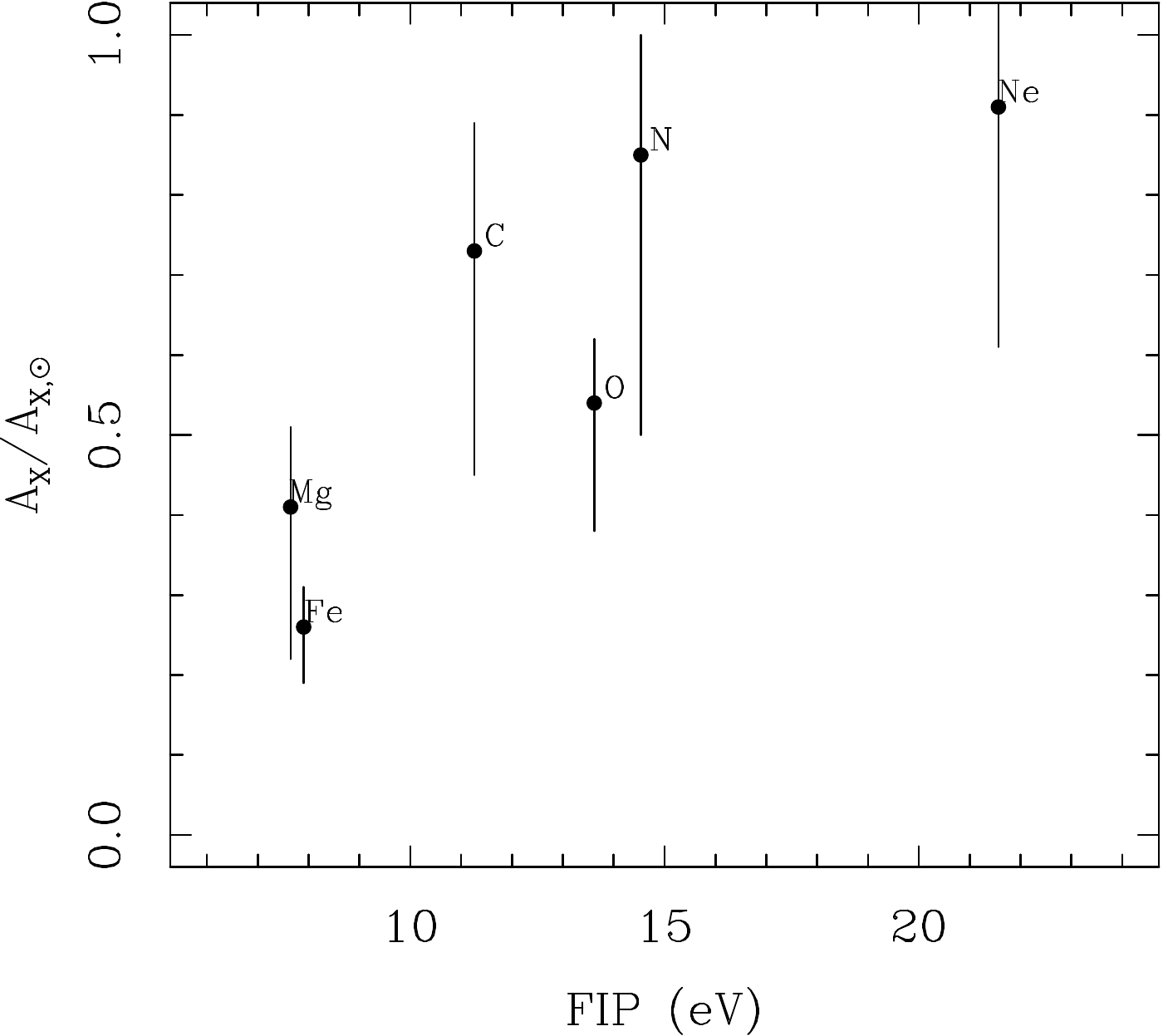}}
    \caption{Ratio $A_X/A_{X\odot}$ vs. first ionization potential derived from the 
    abundances analysis of RGS spectra}
    \label{fig:z_fip}
\end{figure}

The abundances of elements inferred from the line fluxes, scaled to { the values of the solar
photosphere} as a function of First Ionization Potential (FIP), are reported in Fig. \ref{fig:z_fip}. 
Apart from the high uncertainty of the abundance of C, a trend with FIP is visible with Fe, 
Mg and perhaps O being under abundant owing to their low FIP. 
The trend is in agreement with the inverse low FIP effect observed in active stars.


\section{Discussion and conclusions}\label{discussion} 
In this paper we have presented the analysis of
an \xmm\ observation of \dstuc~A which is one of the youngest known planet hosts.  
\dstuc~Ab  is an inflated planet with a radius of about 0.5 R$_\mathrm J$ and with mass less than 
14.4 M$_\oplus$ that will likely shrink to a size of about 2 $R_\oplus$ in the next hundreds 
of Myr \citep{Benatti2021}.
Observations in X-rays are important to understand the processes that can modify both dynamics and chemistry of the
primary atmospheres  of newly born planets. In this context, the present work allows a deep
characterization of the coronal emission and time variability of the star and its effects on 
its close-in planet.  

We detected two bright flares in \dstuc~A that permitted a time-resolved spectroscopy of the two
events.  
The energy released during the first and second flare are about $8\times10^{34}$ erg and
$5\times10^{34}$ erg.  

\citet{Colombo2022} used TESS light curves of \dstuc\ to study the rate of flares of DS Tuc in the optical band 
with an analysis of all available sectors through an iterative method based on Gaussian processes. 
They found that the frequency of flares with energy (in optical band) larger than $2\times10^{32}$ erg amounts to about 2 per day. 
By using the relationship between flare energy released in the optical band and in X-rays inferred by \citet{Flaccomio2018} from the
flares observed in the NGC 2264 young star forming region,  this limit corresponds to X-ray energies of order of  10$^{31}$ erg. 
The flares we detected in the present \xmm\ observation seem very rare events that likely released energies in the optical band in excess of 
$5\times10^{35}$ erg which are on the very high energy tail of the flare energies observed with TESS by Colombo et al (2022).
Furthermore, in the optical band there is a significant fraction of pairs of  flares separated by a few ks,
 similar to the separation of the two flares we   observed in X-rays. The cause could be a triggering mechanism of the same 
active region after the first flare has been ignited in a scenario where the corona is densely packed
with magnetic structures somehow interacting with each other.
The UV energy of the flares was calculated from the count rates and the conversion factor for the UVM2 filter 
(assuming a band width of 100 \AA) in the case of a WD at 10,000 K
\footnote{See \url{https://www.cosmos.esa.int/web/xmm-newton/sas-watchout-uvflux}.}. 
These energies amount to 2.7 and $0.9 \times10^{33}$ erg for flares 1 and 2, 
respectively, and result a factor 30 to 50 lower than the energies released in X-rays.

In pre Main Sequence stars of Orion, { X-ray} flares that release energies  $\ge 5.5\times10^{35}$ erg 
constitute the upper 50\% of the sample of the bright flares observed in COUP \citep[cf. Fig. 9 in ][]{Wolk2005}.  
It is plausible that the frequency of such energetic events decreases between the age of young  
pre-Main Sequence stars in Orion ($3-5$ Myr) and the age of \dstuc, about 40 Myr.

With an age of 40 Myr the X-ray activity of \dstuc~A has just started its decline in the Main Sequence phase
and  probed by a value of the ratio  $\log L_\mathrm X/L_\mathrm{bol}\sim-3.55$ \citep{Benatti2021}.
The average X-ray luminosity of \dstuc~A in this observation was $9.1\times10^{30}$ \lxu which is
about a factor 10 higher than recorded in the first \xmm\ observation \citep{Benatti2021}. 
During the flares the ratio $\log L_\mathrm X/L_\mathrm{bol}$ reached the value of  $\approx-2.15$  
or $\approx0.7\%$ of its bolometric luminosity.

From the ephemeris of \citet{Benatti2019} we estimated that the planet was at the orbital phases
{ $\phi$ between $\sim -0.06$  and  $\sim -0.11$ } during this observation (being $\phi=0$ 
the phase of the transit { and $\sim -0.008$ the phase of the first contact}) 
and thus the planet was  observed between 19 and 12 hours before the transit. 
The stellar rotation period is about 2.9 days which amounts to $\approx250.5$ ks, thus 
the present \xmm\ exposure covered about 12\% of the rotation period. 
The flares did not show any dimming due to self-eclipse, we infer that either they occurred
in active regions near the equator and in the central portion of the stellar disk or 
the active regions were at very high stellar latitudes and coronal 
loops with sizes similar to the stellar radius anchored at high stellar latitudes
could be visible even when the feet of the loops are behind the visible disk.

The values of peak luminosity are about $1.8\times10^{31}$ erg/s, these values
correspond to a flux of about $100$ \fxu at the surface of the planet, 
which is of order of
0.1 W m$^{-2}$. Conversely, the X-ray flux received by the planet from the quiescent star is about 0.01 W m$^{-2}$.
An hypothetical Earth around \dstuc~A at 1 AU would have received $6.3\times10^{-4}$ W m$^{-2}$ in
the band $0.3-10$ keV, which corresponds to a GOES X6 flare in the band $1-8$ \AA.  
We speculate that Coronal Mass Ejections (CMEs) were associated with those flares.
\citet{Drake2013} \citep[see also][]{Osten2015, Moschou2019} give an estimate of the mass and kinetic energy associated with powerful flares
observed in the Sun. From their equation 1, which applies to young stellar 
coronae in the saturated regime like in the case of \dstuc~A, 
we estimated that the mass associated with CMEs was around  $5\times 10^{-15}$ $M_\odot$.  
We also infer that the kinetic energy associated with the CMEs is $\approx10^{34}$ erg 
From these numbers we also infer a velocity of the CME material of order of 1000 km/s. 
With these velocities the CME would have {reached  \dstuc~Ab}  in about $3.3$ hr after 
the ignition of the flares. 
Simulations of CMEs hitting close-in planets at a few stellar radii from their stars
\citep[see][]{Alvarado2022,Hazra2022} demonstrate that
the CMEs associated with the flares could be even more disruptive for the outer atmosphere resulting in
dramatic changes of the rate of evaporation with respect to the effect of a steady stellar wind or
from X-ray irradiation alone due to a flare.

The RGS spectra enable us to reconstruct the distribution of Emission Measure (EMD) for the quiescent
and flaring states. The EMDs are very similar in the range $6 < \log T <7.1$ while they  differ
significantly above $\log T> 7.1$ with the maximum difference occurring at $\log T\approx 7.5$. The 
difference is due to the dense plasma heated during the flares.
The integral of the difference between the flare and quiescent phase EMDs 
in the range of temperature $7 < \log T < 8$ amounts to $\log EM \approx 53.5 $ which is consistent, 
on average,  with the values inferred from the time resolved spectroscopy of EPIC spectra.
Qualitatively, hard X-rays produced during the flares can penetrate more deeply in the planet's atmosphere
along with a cascade of electrons that induce further dissociation and photo-chemistry. 
Such a mechanism can be non linear if the atmosphere is unable to restore its state before the next flare.
\citet{Louca2022} modeled the response of several planetary atmospheres to flares determining the 
conditions under which the initial conditions are restored and the consequent enhancement 
of several species.
In this respect, a more systematic X-ray monitoring of stars with close in planets would be beneficial
for the comprehension of the formation and evolution of planetary atmospheres.

\begin{acknowledgements} 
{ We thank the anonymous referee for their useful comments and suggestions that improved 
the quality of this manuscript.}
IP, AM, SC and GM acknowledge financial support from the ASI-INAF agreement n.2018-16-HH.0, 
and from the ARIEL ASI-INAF agreement n.2021-5-HH.0. 
 SJW\ was supported by the Chandra X-ray Observatory Center, which is operated by the Smithsonian 
 Astrophysical Observatory for and on behalf of the National Aeronautics Space Administration 
 under contract NAS8-03060.
Based on observations obtained with XMM-Newton, an ESA science mission with instruments 
and contributions directly funded by ESA Member States and NASA.

\end{acknowledgements}


\begin{thebibliography}{36}
\expandafter\ifx\csname natexlab\endcsname\relax\def\natexlab#1{#1}\fi

\bibitem[{{Alvarado-G{\'o}mez} {et~al.}(2022){Alvarado-G{\'o}mez}, {Cohen},
  {Drake}, {Fraschetti}, {Poppenhaeger}, {Garraffo}, {Chebly}, {Ilin},
  {Harbach}, \& {Kochukhov}}]{Alvarado2022}
{Alvarado-G{\'o}mez}, J.~D., {Cohen}, O., {Drake}, J.~J., {et~al.} 2022, \apj,
  928, 147

\bibitem[{{Argiroffi} {et~al.}(2019){Argiroffi}, {Reale}, {Drake},
  {Ciaravella}, {Testa}, {Bonito}, {Miceli}, {Orlando}, \&
  {Peres}}]{Argiroffi2019}
{Argiroffi}, C., {Reale}, F., {Drake}, J.~J., {et~al.} 2019, Nature Astronomy,
  3, 742

\bibitem[{{Arnaud}(1996)}]{Arnaud1996}
{Arnaud}, K.~A. 1996, in Astronomical Society of the Pacific Conference Series,
  Vol. 101, Astronomical Data Analysis Software and Systems V, ed. G.~H.
  {Jacoby} \& J.~{Barnes}, 17

\bibitem[{{Benatti} {et~al.}(2021){Benatti}, {Damasso}, {Borsa}, {Locci},
  {Pillitteri}, {Desidera}, {Maggio}, {Micela}, {Wolk}, {Claudi}, {Malavolta},
  \& {Modirrousta-Galian}}]{Benatti2021}
{Benatti}, S., {Damasso}, M., {Borsa}, F., {et~al.} 2021, \aap, 650, A66

\bibitem[{{Benatti} {et~al.}(2019){Benatti}, {Nardiello}, {Malavolta},
  {Desidera}, {Borsato}, {Nascimbeni}, {Damasso}, {D'Orazi}, {Mesa}, {Messina},
  {Esposito}, {Bignamini}, {Claudi}, {Covino}, {Lovis}, \&
  {Sabotta}}]{Benatti2019}
{Benatti}, S., {Nardiello}, D., {Malavolta}, L., {et~al.} 2019, \aap, 630, A81

\bibitem[{{Cecchi-Pestellini} {et~al.}(2009){Cecchi-Pestellini}, {Ciaravella},
  {Micela}, \& {Penz}}]{Cecchi-Pestellini2009}
{Cecchi-Pestellini}, C., {Ciaravella}, A., {Micela}, G., \& {Penz}, T. 2009,
  \aap, 496, 863

\bibitem[{{Colombo} {et~al.}(2022){Colombo}, {Petralia}, \&
  {Micela}}]{Colombo2022}
{Colombo}, S., {Petralia}, A., \& {Micela}, G. 2022, arXiv e-prints,
  arXiv:2202.13615

\bibitem[{{Dennis} \& {Zarro}(1993)}]{Dennis1993}
{Dennis}, B.~R. \& {Zarro}, D.~M. 1993, \solphys, 146, 177

\bibitem[{{Dere} {et~al.}(1997){Dere}, {Landi}, {Mason}, {Monsignori Fossi}, \&
  {Young}}]{Dere1997}
{Dere}, K.~P., {Landi}, E., {Mason}, H.~E., {Monsignori Fossi}, B.~C., \&
  {Young}, P.~R. 1997, \aaps, 125, 149

\bibitem[{{Drake} {et~al.}(2013){Drake}, {Cohen}, {Yashiro}, \&
  {Gopalswamy}}]{Drake2013}
{Drake}, J.~J., {Cohen}, O., {Yashiro}, S., \& {Gopalswamy}, N. 2013, \apj,
  764, 170

\bibitem[{{Favata} {et~al.}(2003){Favata}, {Giardino}, {Micela}, {Sciortino},
  \& {Damiani}}]{Favata03}
{Favata}, F., {Giardino}, G., {Micela}, G., {Sciortino}, S., \& {Damiani}, F.
  2003, \aap, 403, 187

\bibitem[{{Feigelson} \& {Montmerle}(1999)}]{Feigelson99}
{Feigelson}, E.~D. \& {Montmerle}, T. 1999, \araa, 37, 363

\bibitem[{{Flaccomio} {et~al.}(2018){Flaccomio}, {Micela}, {Sciortino}, {Cody},
  {Guarcello}, {Morales-Calder{\`o}n}, {Rebull}, \& {Stauffer}}]{Flaccomio2018}
{Flaccomio}, E., {Micela}, G., {Sciortino}, S., {et~al.} 2018, \aap, 620, A55

\bibitem[{{Hazra} {et~al.}(2022){Hazra}, {Vidotto}, {Carolan}, {Villarreal
  D'Angelo}, \& {Manchester}}]{Hazra2022}
{Hazra}, G., {Vidotto}, A.~A., {Carolan}, S., {Villarreal D'Angelo}, C., \&
  {Manchester}, W. 2022, \mnras, 509, 5858

\bibitem[{{Hudson}(1991)}]{Hudson1991}
{Hudson}, H.~S. 1991, \solphys, 133, 357

\bibitem[{{Kahler} {et~al.}(1988){Kahler}, {Moore}, {Kane}, \&
  {Zirin}}]{Kahler1988}
{Kahler}, S.~W., {Moore}, R.~L., {Kane}, S.~R., \& {Zirin}, H. 1988, \apj, 328,
  824

\bibitem[{{Kashyap} \& {Drake}(1998)}]{Kashyap1998}
{Kashyap}, V. \& {Drake}, J.~J. 1998, \apj, 503, 450

\bibitem[{{Kashyap} \& {Drake}(2000)}]{Kashyap2000}
{Kashyap}, V. \& {Drake}, J.~J. 2000, Bulletin of the Astronomical Society of
  India, 28, 475

\bibitem[{{Kay} {et~al.}(2019){Kay}, {Airapetian}, {L{\"u}ftinger}, \&
  {Kochukhov}}]{Kay2019}
{Kay}, C., {Airapetian}, V.~S., {L{\"u}ftinger}, T., \& {Kochukhov}, O. 2019,
  \apjl, 886, L37

\bibitem[{{Kubyshkina} {et~al.}(2018){Kubyshkina}, {Fossati}, {Erkaev},
  {Cubillos}, {Johnstone}, {Kislyakova}, {Lammer}, {Lendl}, \&
  {Odert}}]{Kubyshkina2018a}
{Kubyshkina}, D., {Fossati}, L., {Erkaev}, N.~V., {et~al.} 2018, \apjl, 866,
  L18

\bibitem[{{Louca} {et~al.}(2022){Louca}, {Miguel}, {Tsai}, {Froning}, {Loyd},
  \& {France}}]{Louca2022}
{Louca}, A.~J., {Miguel}, Y., {Tsai}, S.-M., {et~al.} 2022, arXiv e-prints,
  arXiv:2204.10835

\bibitem[{{Maggio} {et~al.}(2000){Maggio}, {Pallavicini}, {Reale}, \&
  {Tagliaferri}}]{Maggio2000}
{Maggio}, A., {Pallavicini}, R., {Reale}, F., \& {Tagliaferri}, G. 2000, \aap,
  356, 627

\bibitem[{{Mauche} {et~al.}(2001){Mauche}, {Liedahl}, \&
  {Fournier}}]{Mauche2001}
{Mauche}, C.~W., {Liedahl}, D.~A., \& {Fournier}, K.~B. 2001, \apj, 560, 992

\bibitem[{{Moschou} {et~al.}(2019){Moschou}, {Drake}, {Cohen},
  {Alvarado-G{\'o}mez}, {Garraffo}, \& {Fraschetti}}]{Moschou2019}
{Moschou}, S.-P., {Drake}, J.~J., {Cohen}, O., {et~al.} 2019, \apj, 877, 105

\bibitem[{{Namekata} {et~al.}(2017){Namekata}, {Sakaue}, {Watanabe}, {Asai},
  {Maehara}, {Notsu}, {Notsu}, {Honda}, {Ishii}, {Ikuta}, {Nogami}, \&
  {Shibata}}]{Namekata2017}
{Namekata}, K., {Sakaue}, T., {Watanabe}, K., {et~al.} 2017, \apj, 851, 91

\bibitem[{{Neupert}(1968)}]{Neupert1968}
{Neupert}, W.~M. 1968, \apjl, 153, L59

\bibitem[{{Newton} {et~al.}(2019){Newton}, {Mann}, {Tofflemire}, {Pearce},
  {Rizzuto}, {Vanderburg}, {Martinez}, {Wang}, {Ruffio}, {Kraus}, {Johnson},
  {Thao}, {Wood}, {Rampalli}, {Nielsen}, {Collins}, {Dragomir}, {Hellier},
  {Anderson}, {Barclay}, {Brown}, {Feiden}, {Hart}, {Isopi}, {Kielkopf},
  {Mallia}, {Nelson}, {Rodriguez}, {Stockdale}, {Waite}, {Wright}, {Lissauer},
  {Ricker}, {Vanderspek}, {Latham}, {Seager}, {Winn}, {Jenkins}, {Bouma},
  {Burke}, {Davies}, {Fausnaugh}, {Li}, {Morris}, {Mukai}, {Villase{\~n}or},
  {Villeneuva}, {De Rosa}, {Macintosh}, {Mengel}, {Okumura}, \&
  {Wittenmyer}}]{Newton2019}
{Newton}, E.~R., {Mann}, A.~W., {Tofflemire}, B.~M., {et~al.} 2019, \apjl, 880,
  L17

\bibitem[{{Osten} \& {Wolk}(2015)}]{Osten2015}
{Osten}, R.~A. \& {Wolk}, S.~J. 2015, \apj, 809, 79

\bibitem[{{Pallavicini} {et~al.}(1981){Pallavicini}, {Golub}, {Rosner},
  {Vaiana}, {Ayres}, \& {Linsky}}]{Pallavicini81}
{Pallavicini}, R., {Golub}, L., {Rosner}, R., {et~al.} 1981, \apj, 248, 279

\bibitem[{{Penz} \& {Micela}(2008)}]{Penz08}
{Penz}, T. \& {Micela}, G. 2008, \aap, 479, 579

\bibitem[{{Pizzolato} {et~al.}(2003){Pizzolato}, {Maggio}, {Micela},
  {Sciortino}, \& {Ventura}}]{Pizzolato2003}
{Pizzolato}, N., {Maggio}, A., {Micela}, G., {Sciortino}, S., \& {Ventura}, P.
  2003, \aap, 397, 147

\bibitem[{{Reale}(2007)}]{Reale2007}
{Reale}, F. 2007, \aap, 471, 271

\bibitem[{{Reale}(2014)}]{Reale2014}
{Reale}, F. 2014, Living Reviews in Solar Physics, 11, 4

\bibitem[{{Vaiana} {et~al.}(1981){Vaiana}, {Fabbiano}, {Giacconi}, {Golub},
  {Gorenstein}, {Harnden}, {Cassinelli}, {Haisch}, {Johnson}, {Linsky},
  {Maxson}, {Mewe}, {Rosner}, {Seward}, {Topka}, \& {Zwaan}}]{Vaiana1981}
{Vaiana}, G.~S., {Fabbiano}, G., {Giacconi}, R., {et~al.} 1981, \apj, 245, 163

\bibitem[{{Veronig} {et~al.}(2002){Veronig}, {Temmer}, {Hanslmeier}, {Otruba},
  \& {Messerotti}}]{Veronig2002}
{Veronig}, A., {Temmer}, M., {Hanslmeier}, A., {Otruba}, W., \& {Messerotti},
  M. 2002, \aap, 382, 1070

\bibitem[{{Wolk} {et~al.}(2005){Wolk}, {Harnden}, {Flaccomio}, {Micela},
  {Favata}, {Shang}, \& {Feigelson}}]{Wolk2005}
{Wolk}, S.~J., {Harnden}, Jr., F.~R., {Flaccomio}, E., {et~al.} 2005, \apjs,
  160, 423

\end{thebibliography}

\end{document}